# Likelihood scan of the Super-Kamiokande I time series data

Gioacchino Ranucci
*Istituto Nazionale di Fisica Nucleare*
*Via Celoria 16 - 20133 Milano*
*e-mail: gioacchino.ranucci@mi.infn.it*

ABSTRACT

In this work a detailed spectral analysis of the time series of the $^8$B solar neutrino flux published by the Super-Kamiokande Collaboration is presented, performed through a likelihood scan approach. Preliminarily a careful review of the analysis methodology is given, showing that the traditional periodicity search via the Lomb-Scargle periodogram is a special case of a more general likelihood based method. Since the data are published together with the relevant asymmetric errors, it is then shown how the likelihood analysis can be performed either with or without a prior error averaging. A key point of this work is the detailed illustration of the mathematical model describing the statistical properties of the estimated spectra obtained in the various cases, which is also validated through extensive Monte Carlo computations; the model includes a calculation for the prediction of the possible alias effects. In the successive investigation of the data, such a model is used to derive objective, mathematical predictions which are quantitatively compared with the features observed in the experimental spectra. This article clearly demonstrates that the handling of the errors is the origin of the discrepancy between published null observations and claimed significant periodicity in the same SK-I data sample. Moreover, the comprehensive likelihood analysis with asymmetric errors developed in this work provides results which cannot exclude the null hypothesis of constant rate, even though some indications stemming from the model at odd with such conclusion point towards the desirability of additional investigations with alternative methods to shed further light on the characteristics of the data.

# I. INTRODUCTION

Recently a number of periodicity searches on the solar neutrino time series data released by the Super-Kamiokande collaboration have been published, either using the standard Lomb-Scargle periodogram method [1] or a likelihood based methodology [2][3][4]. Both methods are intended to produce an estimate of the power spectrum of the series, with the purpose to unravel periodicities hidden in the noise affecting the data, which would appear as sharp peaks in the spectrum itself. The difficulty associated with such an analysis is that the noise affecting the data points produces in the spectrum random peaks that can attain very high levels, thus hampering the capability to detect properly actual modulations embedded in the series. The crucial aspect of the analysis is thus the significance assessment of the peaks (in particular the largest) found in the estimated spectrum: such a significance is defined as the probability that a peak as high or higher than the highest peak found in the actual spectrum can be generated by chance noise fluctuations. The outcomes of the significance assessment in the published analysis are controversial, with different analysis producing different results.

Purpose of this work is to present a thorough model for the time series data analysis and interpretation, from which derive mathematical predictions that could be extensively compared with the Super-Kamiokande data, highlighting the role of the elements at the origin of the mentioned discrepancy in the published results. To this end the analysis methodologies under consideration (periodogram and likelihood) are first carefully reviewed, showing the close relationship between them, stemming from the fact that they share the same basic approach to the problem of power spectrum estimation. Furthermore, a thorough account is given of the statistical properties of the estimated spectra in the various approaches, showing how to extend the statistical description normally given to the highest spectral peak to the peaks of less height; afterwards, such methodologies are systematically applied to the 10 day and 5 day binned time series of the solar neutrino data published by Super-Kamiokande collaboration.

Specifically, the paper is organized as follow: in the paragraph 2 the basic definition used to derive the Lomb-Scargle periodogram [5] is revisited, and it is demonstrated that it coincides with the definition used to obtain the spectrum stemming from the likelihood methodology, the difference between the two is simply that in the former the data points are considered affected by a common error, while in the latter are affected by unequal individual errors. In the paragraph 3 the definition of the likelihood spectrum is developed to an explicit formulation following a procedure that resembles the least square sinusoidal fit to the data used in [5]. At the end of the paragraph it is pointed out that, through a numerical iterative procedure, the likelihood spectrum can be also computed in the special situation of asymmetric errors affecting the data points. In the paragraph 4, as further step under the assumption of ignoring the error asymmetry, the same numerical approximation adopted still in [5] to get the final form of the periodogram is exploited to obtain from the likelihood spectrum a generalized "weighted" periodogram that accounts for the unequal errors; along such a derivation it is shown that numerically the likelihood spectrum and the generalized weighted periodogram practically coincide (in the case of symmetric errors), thus allowing to replace for practical purposes the former with the latter. In the paragraph 5, as check of the overall procedure, it is then illustrated how from the generalized weighted periodogram the standard Lomb Scargle periodogram is recovered if the data points errors are equal. For completeness it also reported the demonstration given in [6] that, for evenly sampled time series, the standard periodogram reduces in turns to the so called Schuster periodogram, i.e. a power spectrum estimate which stems directly from the application of the Fourier transform to time series sampled at equally spaced points.

In the paragraph 6 the crucial issues of the statistics of the estimated spectrum is addressed: first the specific demonstration that the generalized weighted periodogram shares the same distribution properties of the standard Lomb Scargle periodogram, under the null hypothesis of no periodicity embedded in the series, is given; then resorting to the Wilks' theorem such a property is put in a general context that encompasses all the spectrum definitions being considered. By ordering according to their rank the peaks in the estimated spectrum (the highest, the 2$^{nd}$ highest and so on), it is also introduced an analytical model for the probability density functions of the height of the peaks so ordered, thus extending the statistical treatment normally restricted only to the highest spectral peak. Furthermore, based on the work illustrated in [7], it is presented the model of the expected spectral response at the frequency corresponding to a true periodicity embedded in the series and it is shown how it can be used to predict the location of the alias frequencies.

In paragraph 7 it is simulated the analysis of fictitious time series comprising 100 points either evenly or unevenly sampled, with special emphasis on the Monte Carlo procedure to compute the null hypothesis distributions used to asses the peaks significance. A careful account of the comparison of the Monte Carlo outputs with the statistical model of the previous paragraph is given, as well. In paragraph 8, instead, it is exemplified the expected model output for a true periodicity, performing also in this case a detailed Monte Carlo-model comparison.

In paragraphs 9 and 10 the Super-Kamiokande 10 day binned dataset is extensively analyzed using the three spectral methodologies introduced in the first part of the paper: Lomb-Scargle, weighted periodogram and likelihood with asymmetric errors. Specifically in paragraph 9 it is reported the peaks significance assessment via the comparison with the null hypothesis Monte Carlo distributions, while in paragraph 10 it is shown how the actual spectral features compare with the model expectations for true periodicities present in the series, including also the model prediction for the alias phenomenon. In paragraph 11 an overall data-model comparison is discussed.

Similarly, the paragraphs 12 and 13 contain the same evaluations for the Super-Kamiokande 5 day binned dataset; in paragraph 14 the systematic effects associated with the errors handling are discussed, while in paragraph 15 a complete overall review of the data-model comparison is illustrated, putting in a comprehensive unitary framework the results related to both the 10 and 5 day series. Finally, a digression on the systematic evaluation of the detection efficiency in the 5 day case is reported in paragraph 16.

The summary of the obtained results is given in the last paragraph 17.

Readers interested in the analysis output but not desiring to go deeply into the mathematical details may skip the derivations reported in the paragraphs from 2 to 5; it is however recommended to read the paragraph 6 about the statistical properties of the estimated spectrum: this paragraph is truly the cornerstone of the entire paper since it provides the mathematical model for objective and quantitative predictions to be confronted with the actual experimental spectra. Furthermore, the paragraphs from 2 to 8 represent a sort of short review of the unsmoothed spectral analysis of time series (unsmoothed in the sense that no tapering or windowing is applied to the data), and feature hence a general validity, independently from the subsequent application to the SK data.

## II. SPECTRUM ESTIMATE AS REDUCTION IN THE SUM OF SQUARES AND LIKELIHOOD SPECTRUM

For the sake of the present discussion it is useful to adopt as starting point the definition of the spectrum of a series as reduction in the sum of squares, following the definition given in [8] in the framework of astronomical studies. The concept is simple: given a series $x_k$, from which the average value $F$ is preliminarily subtracted, one can construct two sum of squares, that of the $x_k$ $\sum_{k=1}^{N} x_k^2$ and that of the $x_k$ subtracted of a quantity $X_k$ resulting from some fit procedure $\sum_{k=1}^{N} (x_k - X_k)^2$. The reduction in the sum of squares

$$\sum_{k=1}^{N} x_k^2 - \sum_{k=1}^{N} (x_k - X_k)^2 \qquad (1)$$

attains a maximum when the fit is good, because obviously in such a case the subtracted terms are minimized. Starting from this simple formulation, and adopting a least square sinusoidal fit to the data to determine the $X_k$, Lomb derived its periodogram [5].

This straightforward definition can be put in a more general context via a likelihood approach. The fit of the $X_k$ to the data can be considered obtained via a likelihood maximization, i.e.

$$(X_{1mx}, X_{2\max},...., X_{N\max}) \Rightarrow \max L = \max e^{-\frac{1}{2}\left(\sum_{k=1}^{N} \frac{(x_k - X_k)^2}{\sigma_k^2}\right)} \qquad (2)$$

where the $\sigma_k$'s are the errors affecting each measured term $x_k$

The amount of increase of *max L* over the "fit to zero" likelihood

$$LR = \frac{\max e^{-\frac{1}{2}\left(\sum_{k=1}^{N} \frac{(x_k - X_k)^2}{\sigma_k^2}\right)}}{e^{-\frac{1}{2}\sum_{k=1}^{N}\frac{x_K^2}{\sigma_k^2}}} \qquad (3)$$

is higher when the fit is good.

If, instead of using as indicator of the presence of a good model fit to the data the (3), we use its logarithm

$$S = \frac{1}{2}\sum_{k=1}^{N} \frac{x_K^2}{\sigma_k^2} - \frac{1}{2}\min \sum_{k=1}^{N} \frac{(x_k - X_k)^2}{\sigma_k^2} \qquad (4)$$

we re-obtain essentially the (1), apart an inessential factor ½, with in addition the generalization of the inclusion of the errors. (It is worth to point out that in the final formulation of the Lomb-Scargle periodogram Scargle [6] added the factor ½ for normalization purpose, and hence the parallelism between the (4) and the Lomb-Scargle definition is total). Actually, in the (1) the errors were implicitly assumed equal for all the points $\sigma_k = \sigma$ and hence without the need to be explicitly included.

If the $X_k$ in (4) are obtained via a frequency dependent sinusoidal fit to the data, then the (4) itself becomes a frequency dependent function: the likelihood spectrum, i.e. the likelihood estimate of the power spectrum of the original data series.

### III. EXPLICIT FORMULATION OF THE LIKELIHOOD SPECTRUM

Let's now proceed to put the (4) in an explicit form for the above mentioned sinusoidal case. For simplicity we follow the notation of the Lomb paper [5]. Hence, considering that the series has been preliminarily averaged to zero, we fit it with oscillations around zero written as

$$A\cos\omega t + B\sin\omega t \tag{5}$$

hence if the terms of the series $x_k$ feature a non zero average $F$, then they are simply replaced by

$$x_k - F \tag{6}.$$

The likelihood is thus

$$L = e^{-\frac{1}{2}\left(\sum_{k=1}^{N} \frac{[x_k - (A\cos\omega t_k + B\sin\omega t_k)]^2}{\sigma_k^2}\right)} \tag{7}$$

which is maximized when the term

$$E = \sum_{k=1}^{N} \frac{[x_k - (A\cos\omega t_k + B\sin\omega t_k)]^2}{\sigma_k^2} \tag{8}$$

gets the minimum. Furthermore, we can write the likelihood ratio as

$$LR = \frac{\max(A,B) e^{-\frac{1}{2}\left(\sum_{k=1}^{N} \frac{[x_k - (A\cos\omega t_k + B\sin\omega t_k)]^2}{\sigma_k^2}\right)}}{e^{-\frac{1}{2}\sum_{k=1}^{N} \frac{x_k^2}{\sigma_k^2}}} \tag{9}$$

from which the spectrum definition of the previous paragraph becomes

$$S(\omega) = \frac{1}{2}\sum_{k=1}^{N} \frac{x_k^2}{\sigma_k^2} - \min(A,B)\frac{1}{2}\sum_{k=1}^{N} \frac{[x_k - (A\cos\omega t_k + B\sin\omega t_k)]^2}{\sigma_k^2} \tag{10}.$$

In order to accomplish explicitly the minimization operation in (10), we can follow the same procedure used by Lomb, generalized to the current case in which the $\sigma_k$ terms are present.

First we proceed to minimize (8) by performing the partial derivatives

$$\frac{\partial E}{\partial A} = \sum_{k=1}^{N} \frac{2}{\sigma_k^2}[x_k - (A\cos\omega t_k + B\sin\omega t_k)](-\cos\omega t_k) \tag{11}$$

$$\frac{\partial E}{\partial B} = \sum_{k=1}^{N} \frac{2}{\sigma_k^2}[x_k - (A\cos\omega t_k + B\sin\omega t_k)](-\sin\omega t_k) \tag{12}$$

and equating both equations to zero we get

$$A\sum_{k=1}^{N}\frac{\cos^2 \omega t_k}{\sigma_k^2} + B\sum_{k=1}^{N}\frac{\sin \omega t_k \cos \omega t_k}{\sigma_k^2} = \sum_{k=1}^{N}\frac{x_k}{\sigma_k^2}\cos \omega t_k \qquad (13)$$

$$A\sum_{k=1}^{N}\frac{\sin \omega t_k \cos \omega t_k}{\sigma_k^2} + B\sum_{k=1}^{N}\frac{\sin^2 \omega t_k}{\sigma_k^2} = \sum_{k=1}^{N}\frac{x_k}{\sigma_k^2}\sin \omega t_k \qquad (14)$$

which can be written in matrix notation

$$\begin{pmatrix} \sum_{k=1}^{N}\frac{\cos^2 \omega t_k}{\sigma_k^2} & \sum_{k=1}^{N}\frac{\sin \omega t_k \cos \omega t_k}{\sigma_k^2} \\ \sum_{k=1}^{N}\frac{\sin \omega t_k \cos \omega t_k}{\sigma_k^2} & \sum_{k=1}^{N}\frac{\sin^2 \omega t_k}{\sigma_k^2} \end{pmatrix} \begin{pmatrix} A \\ B \end{pmatrix} = \begin{pmatrix} \sum_{k=1}^{N}\frac{x_k}{\sigma_k^2}\cos \omega t_k \\ \sum_{k=1}^{N}\frac{x_k}{\sigma_k^2}\sin \omega t_k \end{pmatrix} \qquad (15).$$

By defining

$$cc = \sum_{k=1}^{N}\frac{\cos^2 \omega t_k}{\sigma_k^2} \quad cs = \sum_{k=1}^{N}\frac{\sin \omega t_k \cos \omega t_k}{\sigma_k^2} \quad ss = \sum_{k=1}^{N}\frac{\sin^2 \omega t_k}{\sigma_k^2} \qquad (16)$$

$$\Delta = \begin{pmatrix} cc & cs \\ cs & ss \end{pmatrix} \qquad (17)$$

we get finally

$$\begin{pmatrix} A \\ B \end{pmatrix} = \Delta^{-1} \begin{pmatrix} \sum_{k=1}^{N}\frac{x_k}{\sigma_k^2}\cos \omega t_k \\ \sum_{k=1}^{N}\frac{x_k}{\sigma_k^2}\sin \omega t_k \end{pmatrix} \qquad (18)$$

being

$$\Delta^{-1} = \begin{pmatrix} \frac{ss}{D} & -\frac{cs}{D} \\ -\frac{cs}{D} & \frac{ss}{D} \end{pmatrix} \qquad D = cc \cdot ss - cs^2 \qquad (19).$$

At this point substituting the (18) in the (10) we get the explicit and exact formulation of the likelihood spectrum.

It must be pointed out that the result obtained in this way is fully equivalent to the alternative approach of simply performing the minimization in (10) by some numerical iterative procedure. In this case it is customary to write the trial sinusoidal function, instead as the (5), as

$$AF \sin(\omega t_k + \varphi) \qquad (20)$$

$A$ being thus the amplitude of the oscillation, expressed as fraction of the average value $F$, and $\varphi$ being the phase; consistently the spectrum is written as

$$S(\omega) = \frac{1}{2}\sum_{k=1}^{N} \frac{x_k^2}{\sigma_k^2} - \min(A,\varphi)\frac{1}{2}\sum_{k=1}^{N} \frac{[x_k - (AF\sin(\omega t_k + \varphi))]^2}{\sigma_k^2} \qquad (21)$$

Even if the numerical procedure has the disadvantage of being much more time consuming than the implementation of the analytical solution, it has the advantage of allowing concurrently the determination of the uncertainty on the fitted amplitude $A$ of the trial modulation function. More important, the numerical iterative procedure applied to the (21) can be generalized to the case in which, instead of having a single $\sigma_k$ for each point of the series, there are a couple of asymmetric errors $\sigma_{kup}$ and $\sigma_{kdown}$, while obviously the above analytical solution cannot be afforded in presence of asymmetric errors.

In particular in case of asymmetric errors the (21) becomes

$$S(\omega) = \frac{1}{2}\sum_{k=1}^{N} \frac{x_k^2}{\sigma_{kp1}^2} - \min(A,\varphi)\frac{1}{2}\sum_{k=1}^{N} \frac{[x_k - (AF\sin(\omega t_k + \varphi))]^2}{\sigma_{kp2}^2} \qquad (22)$$

where

$\sigma_{kp2}=\sigma_{kdown}$ if $x_k > (AF\sin(\omega t_k + \varphi))$ and $\sigma_{kp2}=\sigma_{kup}$ if $x_k < (AF\sin(\omega t_k + \varphi))$

and

$\sigma_{kp1}=\sigma_{kdown}$ if $x_k > 0$ and $\sigma_{kp1}=\sigma_{kup}$ if $x_k < 0$.

## IV. DERIVING THE GENERALIZED WEIGHTED PERIODOGRAM FROM THE LIKELIHOOD SPECTRUM

In [5], in the hypothesis of common $\sigma$, it is shown how to derive the standard Lomb-Scargle periodogram through a suitable approximation of expressions similar to those reported in the previous paragraph. The same treatment can be extended to the present case of unequal $\sigma_k$, leading to a weighted form of the periodogram, which automatically allows the proper inclusions of the errors. To this purpose we can manipulate the (10) itself as follows (omitting for simplicity the indication *min*)

$$\frac{1}{2}\sum_{k=1}^{N}\frac{x_k^2}{\sigma_k^2} - \frac{1}{2}\sum_{k=1}^{N}\frac{x_k^2}{\sigma_k^2} - \frac{1}{2}\sum_{k=1}^{N}\frac{(A\cos\omega t_k + B\sin\omega t_k)^2}{\sigma_k^2} + \sum_{k=1}^{N}\frac{x_k(A\cos\omega t_k + B\sin\omega t_k)}{\sigma_k^2} \qquad (23)$$

which becomes

$$\sum_{k=1}^{N}\frac{x_k(A\cos\omega t_k + B\sin\omega t_k)}{\sigma_k^2} - \frac{1}{2}\sum_{k=1}^{N}\frac{(A\cos\omega t_k + B\sin\omega t_k)^2}{\sigma_k^2} \qquad (24)$$

and

$$\sum_{k=1}^{N} \frac{[2x_k - (A\cos\omega t_k + B\sin\omega t_k)](A\cos\omega t_k + B\sin\omega t_k)}{2\sigma_k^2} \tag{25}$$

By writing the last expression as

$$\sum_{k=1}^{N} \frac{\{x_k + [x_k - (A\cos\omega t_k + B\sin\omega t_k)]\}(A\cos\omega t_k + B\sin\omega t_k)}{2\sigma_k^2} \tag{26}$$

it is easily recognized that the term in brackets $[x_k - (A\cos\omega t_k + B\sin\omega t_k)]$, being the residual of the fit, is negligible with respect to $x_k$. The (26) thus becomes approximately equal to

$$\sum_{k=1}^{N} \frac{x_k(A\cos\omega t_k + B\sin\omega t_k)}{2\sigma_k^2} \tag{27}$$

that in matrix form can be written

$$\frac{1}{2}\left(\sum_{k=1}^{N} \frac{x_k \cos\omega t_k}{\sigma_k^2} \quad \sum_{k=1}^{N} \frac{x_k \sin\omega t_k}{\sigma_k^2}\right)\binom{A}{B} \tag{28}$$

or, according to (18) and (19)

$$\frac{1}{2}\left(\sum_{k=1}^{N} \frac{x_k \cos\omega t_k}{\sigma_k^2} \quad \sum_{k=1}^{N} \frac{x_k \sin\omega t_k}{\sigma_k^2}\right)\begin{pmatrix} \frac{ss}{D} & -\frac{cs}{D} \\ -\frac{cs}{D} & \frac{cc}{D} \end{pmatrix}\begin{pmatrix} \sum_{k=1}^{N} \frac{x_k \cos\omega t_k}{\sigma_k^2} \\ \sum_{k=1}^{N} \frac{x_k \sin\omega t_k}{\sigma_k^2} \end{pmatrix} \tag{29}$$

Following Lomb, the 2x2 matrix in (29) can be put in diagonal form making $cs=0$. As shown in [5], this is obtained inserting a time shift such that the (5) becomes

$$A\cos\omega(t-\tau) + B\sin\omega(t-\tau) \tag{30}$$

where $\tau$ is deduced from the equation

$$\frac{\sum_{k=1}^{N} \frac{\sin 2\omega t_k}{\sigma_k^2}}{\sum_{k=1}^{N} \frac{\cos 2\omega t_k}{\sigma_k^2}} = \tan 2\omega\tau \tag{31}$$

With this choice the (27) becomes

$$\frac{1}{2}\left(\sum_{k=1}^{N}\frac{x_k \cos\omega(t_k-\tau)}{\sigma_k^2} \quad \sum_{k=1}^{N}\frac{x_k \sin\omega(t_k-\tau)}{\sigma_k^2}\right)\begin{pmatrix}\frac{1}{cc} & 0 \\ 0 & \frac{1}{ss}\end{pmatrix}\begin{pmatrix}\sum_{k=1}^{N}\frac{x_k \cos\omega(t_k-\tau)}{\sigma_k^2} \\ \sum_{k=1}^{N}\frac{x_k \sin\omega(t_k-\tau)}{\sigma_k^2}\end{pmatrix} \quad (32)$$

which can be manipulated to obtain

$$\frac{1}{2}\left(\frac{1}{cc}\sum_{k=1}^{N}\frac{x_k \cos\omega(t_k-\tau)}{\sigma_k^2} \quad \frac{1}{ss}\sum_{k=1}^{N}\frac{x_k \sin\omega(t_k-\tau)}{\sigma_k^2}\right)\begin{pmatrix}\sum_{k=1}^{N}\frac{x_k \cos\omega(t_k-\tau)}{\sigma_k^2} \\ \sum_{k=1}^{N}\frac{x_k \sin\omega(t_k-\tau)}{\sigma_k^2}\end{pmatrix} \quad (33)$$

$$\frac{1}{2}\left[\frac{1}{cc}\left(\sum_{k=1}^{N}\frac{x_k \cos\omega(t_k-\tau)}{\sigma_k^2}\right)^2 + \frac{1}{ss}\left(\sum_{k=1}^{N}\frac{x_k \sin\omega(t_k-\tau)}{\sigma_k^2}\right)^2\right] \quad (34)$$

and finally, remembering (16)

$$\frac{1}{2}\frac{\left(\sum_{k=1}^{N}\frac{x_k \cos\omega(t_k-\tau)}{\sigma_k^2}\right)^2}{\sum_{k=1}^{N}\frac{\cos^2\omega(t_k-\tau)}{\sigma_k^2}} + \frac{1}{2}\frac{\left(\sum_{k=1}^{N}\frac{x_k \sin\omega(t_k-\tau)}{\sigma_k^2}\right)^2}{\sum_{k=1}^{N}\frac{\sin^2\omega(t_k-\tau)}{\sigma_k^2}} \quad (35)$$

(we remind that for a series with not zero $F$ average $x_k$ is to be considered replaced by $x_k - F$).

The (35) is the weighted periodogram that generalizes the Lomb-Scargle periodogram to take into account the errors associated with each data point. From the relevant derivation, it is clear that such a weighted periodogram represents a numerical approximation of the likelihood spectrum. Actually, it is a very good approximation, as it was checked computing with the two different methods the spectra of the Super-Kamiokande data, and observing an agreement of the spectral values up to the second, or third, decimal digit. It should be noted that the weighted periodogram (35) is different from the weighted form of the Lomb-Scargle periodogram introduced in [9], and used in [3] as an intermediate step for the Super-Kamiokande data analysis.

### V. FROM THE WEIGTHED PERIODOGRAM TO THE STANDARD PERIODOGRAMS

These last two steps are simpler. From the (35), if all the $\sigma_k$ are equal, we get immediately

$$\frac{1}{2\sigma^2}\left(\frac{\left(\sum_{k=1}^{N}x_k \cos\omega(t_k-\tau)\right)^2}{\sum_{k=1}^{N}\cos^2\omega(t_k-\tau)} + \frac{\left(\sum_{k=1}^{N}x_k \sin\omega(t_k-\tau)\right)^2}{\sum_{k=1}^{N}\sin^2\omega(t_k-\tau)}\right) \quad (36)$$

with $\tau$ given by

$$\frac{\sum_{k=1}^{N} \sin 2\omega t_k}{\sum_{k=1}^{N} \cos 2\omega t_k} = \tan 2\omega\tau \qquad (37)$$

which is the well known formulation of the standard Lomb-Scargle periodogram.

There is, however, an important conceptual difference between the (35) and the (36) concerning the errors: while the former implies that the errors are individually known, in the sense that each data point is given together with the relevant error estimate, the (36) is normally interpreted with σ unknown, which hence must be derived from the scatter of the measured values, i.e.

$$\sigma^2 = \frac{1}{N-1}\sum_{k=1}^{N} x_k^2 \qquad (38).$$

Finally, the transformation of the Lomb-Scargle to the normal Schuster periodogram in case of even sampling is well known [6]: the standard Schuster periodogram is evaluated only at a finite set of frequencies $\frac{2\pi}{T}j$ with $j$ from $1$ to $N/2$ (the case j=0 is not included because the average value is preliminarily subtracted); $T$ is the total time interval, $N$ is the number of sampling points and $(T/N)k$ are the sampling points. For the specific set of natural frequencies $\frac{2\pi}{T}j$ ($j$ from $1$ to $N/2$) it can be shown that

$$\sum_{k=1}^{N} \sin 2\omega t_k = 0 \qquad (39)$$

and hence from the (37) it stems that *τ=0*.
Furthermore

$$\sum_{k=1}^{N} \cos^2 \omega(t_k - \tau) = \sum_{k=1}^{N} \cos^2 \frac{2\pi}{N} jk = \frac{N}{2} \qquad (40)$$

and

$$\sum_{k=1}^{N} \sin^2 \omega(t_k - \tau) = \sum_{k=1}^{N} \sin^2 \frac{2\pi}{N} jk = \frac{N}{2} \qquad (41)$$

The (36) thus becomes

$$\frac{1}{N\sigma^2}\left[\left(\sum_{k=1}^{N} x_k \cos \frac{2\pi}{N} jk\right)^2 + \left(\sum_{k=1}^{N} x_k \sin \frac{2\pi}{N} jk\right)^2\right] \qquad (42)$$

which is indeed the Schuster periodogram [10] for even sampling, as derived from the direct application of the Fourier transform to an evenly sampled series.

# VI. STATISTICAL PROPERTIES OF PERIODOGRAMS

It is well known that the ordinate of each frequency of the Schuster periodogram of a purely noisy time series features an exponential distribution [7]. The same property for the standard Lomb Scargle periodogram has been proved in [6] (see also [14]).

Also the generalized weighted periodogram (35) features the same distribution property, as demonstrated in this paragraph. Indeed, following [6], in the case of purely noisy series (i.e. no modulation embedded) the (35) can be considered the sum of the squares of two normally distributed zero mean random variables, i.e.

$$\left( \frac{\sum_{k=1}^{N} \frac{x_k \cos \omega(t_k - \tau)}{\sigma_k^2}}{\sqrt{2 \sum_{k=1}^{N} \frac{\cos^2 \omega(t_k - \tau)}{\sigma_k^2}}} \right)^2 + \left( \frac{\sum_{k=1}^{N} \frac{x_k \sin \omega(t_k - \tau)}{\sigma_k^2}}{\sqrt{2 \sum_{k=1}^{N} \frac{\sin^2 \omega(t_k - \tau)}{\sigma_k^2}}} \right)^2 \qquad (43).$$

The terms in the brackets are linear combinations of the variables $x_k$: since they are zero mean and normally distributed variables, this ensures that the same property is featured also by their combinations.

The term within the first bracket features a variance which is equal to

$$\frac{\sum_k \sum_j \frac{\langle x_k x_j \rangle \cos \omega(t_k - \tau) \cos \omega(t_j - \tau)}{\sigma_k^2 \sigma_j^2}}{2 \sum_{k=1}^{N} \frac{\cos^2 \omega(t_k - \tau)}{\sigma_k^2}} = \frac{1}{2} \qquad (44)$$

(this result is obtained considering that $\langle x_k x_j \rangle = 0$ if $k \neq j$ and $\langle x_k x_j \rangle = \sigma_k^2$ if $k = j$).

Similarly, also the variance of the variable in the second bracket of (43) is equal to ½. The correlation of the two terms is

$$\frac{\sum_k \sum_j \frac{\langle x_k x_j \rangle \cos \omega(t_k - \tau) \sin \omega(t_j - \tau)}{\sigma_k^2 \sigma_j^2}}{2 \sqrt{\sum_{k=1}^{N} \frac{\cos^2 \omega(t_k - \tau)}{\sigma_k^2} \sum_{k=1}^{N} \frac{\sin^2 \omega(t_k - \tau)}{\sigma_k^2}}} = \frac{\sum_k \frac{\cos \omega(t_k - \tau) \sin \omega(t_k - \tau)}{\sigma_k^2}}{2 \sqrt{\sum_{k=1}^{N} \frac{\cos^2 \omega(t_k - \tau)}{\sigma_k^2} \sum_{k=1}^{N} \frac{\sin^2 \omega(t_k - \tau)}{\sigma_k^2}}} = 0 \qquad (45)$$

The equality to zero is due to the fact that the numerator is the *cs* factor (16), forced to be zero with the suitable choice of $\tau$.

It can now be used the result that the distribution of the sum of the squares of two independent gaussian random variables with zero mean and equal variance $\sigma$ is [6]

$$\frac{1}{2\sigma^2} e^{-\frac{z}{2\sigma^2}} \qquad (46)$$

and, since in the present case $\sigma^2$ is ½, we get the result that the distribution of the modified Lomb-Scargle periodogram is simply $e^{-z}$, like that of the standard Lomb-Scargle periodogram [6] (it is worth to repeat this result is valid if there is no modulation in the time series).

It can be shown that the exponential distribution property shared by the Schuster periodogram, the Lomb-Scargle periodogram and the newly introduced weigthed periodogram are simply particular manifestations of the log-likelihood ratio theorem (or Wilks' theorem) [11][12]. More generally, such a theorem proves that the starting point of all these spectra, i.e. the likelihood spectrum, shares the same property, as well.

Indeed the Wilks' theorem states that, given N occurrences of the random variable $x$ obeying to the pdf $p(x, \theta_1 \theta_2 .......\theta_t)$ depending upon the $t$ parameters $\theta_1\ \theta_2 ......\theta_t$, and constructed the generalized likelihood ratio

$$GLR = \frac{\max(\theta_{q+1}...\theta_t) \prod_{i=1}^{N} p(x_i, \theta_1...\theta_q \theta_{q+1}...\theta_t)}{\max(\theta_1 \theta_2 .....\theta_t) \prod_{i=1}^{N} p(x_i, \theta_1 \theta_2 ......\theta_t)} \qquad (47)$$

(hence the maximization at the numerator is done keeping fixed the subset of parameters from $\theta_1\ \theta_2\ ......\theta_q$, while the maximization of the denominator is over the full set of parameters) the quantity -2ln(GLR) under the null hypothesis is asymptotically distributed as $\chi^2(q)$, where q is the number of degrees of freedom.
In our case the (47) specializes to

$$GLR = \frac{e^{-\frac{1}{2}\left(\sum_{k=1}^{N} \frac{x_k^2}{\sigma_k^2}\right)}}{\max(A,\varphi) e^{-\frac{1}{2}\left(\sum_{k=1}^{N} \frac{(x_k - AF\sin(\omega t_k + \varphi))^2}{\sigma_k^2}\right)}} \qquad (48)$$

and therefore the quantity -2ln(GLR) is

$$-2\ln(GLR) = \sum_{k=1}^{N} \frac{x_k^2}{\sigma_k^2} - \min(A,\varphi) \sum_{k=1}^{N} \frac{(x_k - AF\sin(\omega t_k + \varphi))^2}{\sigma_k^2} \qquad (49)$$

which, apart a factor 1/2, coincides with the definition of the spectrum in the framework of the likelihood method. The Wilks' theorem thus ensures that this quantity, under the null hypothesis, is asymptotically distributed as $\chi^2(2)$, i.e. as $\frac{1}{2}e^{-\frac{z}{2}}$, and consequently the likelihood spectrum, being simply the (49) multiplied by ½, is asymptotically distributed as $e^{-z}$. Having demonstrated that the weighted periodogram, the standard Lomb-Scarge periodogram and the Schuster periodogram are simply special cases of the likelihood spectrum, this result applies as well to them: we have thus recovered the previous property in a more general context.

The exponential distribution of the Schuster and of the standard Lomb Scargle periodogram is the basis of the false alarm probability formula, described in [6] and [14], to perform the significance assessment of the largest detected peak. This false alarm formula, which is

$$P(> z) = 1 - (1 - e^{-z})^M \qquad (50)$$

where M is the number of independent scanned frequencies, stems from the probability density function of the height of the largest spectral peak in case of pure noise series: if the peaks are M, each individually exponentially distributed, the PDF of the largest is obviously

$$p_{largest}(h) = M\left[1 - e^{-h}\right]^{M-1} e^{-h} \qquad (51)$$

from which the integration above a threshold z gives the (50).

The above demonstration that also the likelihood spectrum and the weighted periodogram share the same exponential distribution allows to extend to them the same false alarm formula (50). For completeness, it is worth to remind that in the case of the Schuster periodogram for even sampled series M is unambiguously defined and is equal to the number of frequencies at which the periodogram is computed, i.e. N/2 so called natural frequencies, N being the number of sampling points (see the clear explanation in [13]). In the case of uneven sampling M is not easily a-priori defined: it depends on the coherence of the series and is heuristically interpreted as the effective number of independent scanned frequencies; actually the correct value of M is determined via Monte Carlo, in the sense that many synthetic data sets with the same properties of the experimental series under study are generated and the Monte Carlo distribution of the largest spectral peak is fitted to the (51) with a suitable choice of M. A thorough account of this procedure is given for example in [18].

It may be useful to extend the (51) to give the probability density function of the height of all the peaks, not only of the largest. By ordering the peaks at the various frequencies over the search band in term of their height, it is in particular possible to express analytically the distribution of the height of the lowest peak, of the second lowest peak and so on, up to the highest peak, in case of pure white noise time series.

The author has already solved such a problem, in the different context of photoelectron statistics, in [15] (an alternative demonstration of the same result has been given in [16]). The result in [15] applied to the present situation states that, if the number of independent frequencies is M, then the probability density function of the spectrum ordinate of the $i_{th}$ (in term of height, starting from the lowest) spectral peak is

$$p_i(h/M) = \frac{M!}{(i-1)!(M-i)!}[1 - F(h)]^{(M-i)}[F(h)]^{i-1} p(h) \qquad (52)$$

where

$$F(h) = \int_0^h p(\lambda)d\lambda \qquad (53)$$

being $p(h)$ simply $e^{-z}$.
In particular from the (52) the distribution of the highest peak is

$$p_M(h/M) = M\left[1 - e^{-h}\right]^{M-1} e^{-h} \qquad (54)$$

which correctly coincides with the (51).

Up to now we have assumed that the series is purely originated by noise. On the other hand, if there is a modulation embedded in it, the distribution of the height of the spectrum at the corresponding frequency is obviously altered. In the following, to address this case, the guidelines reported in [7] are followed. Denoting with $\omega_s$ the signal frequency, the (43), written for that frequency, specializes to

$$\left(\frac{\sum_{k=1}^{N}\frac{S_k\cos\omega_s(t_k-\tau)+e_k\cos\omega_s(t_k-\tau)}{\sigma_k^2}}{\sqrt{2\sum_{k=1}^{N}\frac{\cos^2\omega_s(t_k-\tau)}{\sigma_k^2}}}\right)^2+\left(\frac{\sum_{k=1}^{N}\frac{S_k\sin\omega_s(t_k-\tau)+e_k\sin\omega_s(t_k-\tau)}{\sigma_k^2}}{\sqrt{2\sum_{k=1}^{N}\frac{\sin^2\omega_s(t_k-\tau)}{\sigma_k^2}}}\right)^2 \quad (55)$$

where $S_k$ is

$$S_k=\frac{1}{(t_{e,k}-t_{s,k})}\int_{t_{s,k}}^{t_{e,k}}A\sin(\omega_s t+\varphi)dt \quad (56)$$

since in this case the $x_k$ samples are the sum of a real sinusoidal signal (that, because of the characteristics of the SK data, is integrated, and not simply sampled, over a detection period of which $t_{sk}$ and $t_{ek}$ are the start and end times) plus the noise terms, denoted with $e_k$.
Writing the (55) as

$$\left(\frac{\sum_{k=1}^{N}\frac{S_k\cos\omega_s(t_k-\tau)}{\sigma_k^2}+\sum_{k=1}^{N}\frac{e_k\cos\omega_s(t_k-\tau)}{\sigma_k^2}}{\sqrt{2\sum_{k=1}^{N}\frac{\cos^2\omega_s(t_k-\tau)}{\sigma_k^2}}}\right)^2+\left(\frac{\sum_{k=1}^{N}\frac{S_k\sin\omega_s(t_k-\tau)}{\sigma_k^2}+\sum_{k=1}^{N}\frac{e_k\sin\omega_s(t_k-\tau)}{\sigma_k^2}}{\sqrt{2\sum_{k=1}^{N}\frac{\sin^2\omega_s(t_k-\tau)}{\sigma_k^2}}}\right)^2 \quad (57)$$

the arguments developed to demonstrate the statistical properties of the (43) extend here showing that the two terms in the brackets are independent Gaussian random variables (denoted respectively $X$ and $Y$) with variance equal to $1/2$, and non zero mean value given respectively by

$$X_m=\frac{\sum_{k=1}^{N}\frac{S_k\cos\omega_s(t_k-\tau)}{\sigma_k^2}}{\sqrt{2\sum_{k=1}^{N}\frac{\cos^2\omega_s(t_k-\tau)}{\sigma_k^2}}} \quad (58)$$

and by

$$Y_m=\frac{\sum_{k=1}^{N}\frac{S_k\sin\omega_s(t_k-\tau)}{\sigma_k^2}}{\sqrt{2\sum_{k=1}^{N}\frac{\sin^2\omega_s(t_k-\tau)}{\sigma_k^2}}} \quad (59)$$

The (58) and (59) are valid also in the case of the standard Lomb Scargle periodogram, provided that one puts the $\sigma_k$ factors equal to the common value $\sigma$.

The ordinate of the spectral line at the frequency corresponding to the signal thus is a random variable $Z$ given by the sum of the squares of the two normal variables $X$ and $Y$ with distributions respectively

$$\frac{1}{\sqrt{\pi}} e^{-(X - X_m)^2} \tag{60}$$

and

$$\frac{1}{\sqrt{\pi}} e^{-(Y - Y_m)^2} \tag{61}.$$

It is known that the sum of the square of normal, independent variables produces a random variable with PDF belonging to the family of the chi square functions (in this case the non central chi square functions). However, instead of using the general expression of this family of curves, it is easier to use directly the probabilistic rules for the combination of random variables [17], from which it can be inferred that the PDF of $Z$ can be expressed through a numerical integration as

$$\int_0^{2\pi} \frac{1}{2\pi} e^{-(\sqrt{Z}\cos\theta - X_m)^2 - (\sqrt{Z}\sin\theta - Y_m)^2} d\theta \tag{62}.$$

The (62) will be extensively used to assess the spectral properties when a signal is present. It may be worth to note that when $X_m$ and $Y_m$ are zero (no signal present) the (62) becomes

$$\int_0^{2\pi} \frac{1}{2\pi} e^{-(\sqrt{Z}\cos\theta)^2 - (\sqrt{Z}\sin\theta)^2} d\theta \tag{63}$$

or

$$\int_0^{2\pi} \frac{1}{2\pi} e^{-Z\cos^2\theta - Z\sin^2\theta} d\theta \tag{64}$$

and finally

$$\int_0^{2\pi} \frac{1}{2\pi} e^{-Z} d\theta = e^{-Z} \tag{65}$$

In this way we have, hence, concretely proved the exponential distribution of the periodogram in case of pure noise.

The numerical integral (62) can be used, as well, to perform the prediction of the expected alias frequencies for a true signal embedded in the series. For this purpose it is enough to compute $X_m$ and $Y_m$ (58) and (59) at other frequencies than the frequency $\omega_s$ of the true signal $S_k$, and check if at any of these other frequencies the integration (62) predicts a significant spectral response.

## VII. NULL HYPOTHESIS DISTRIBUTIONS: MONTE CARLO EXAMPLES AND COMPARISON WITH THE MODEL

It is largely reported in the literature concerning time series that the significance assessment of the largest peak found in the periodogram of an experimental data set is normally done via Monte Carlo [18]. In practice, given the characteristics of the time series under examination, i.e. sampling times sequence and noise variance (or variances in case of unequal measurements errors), a large number of fictitious, purely noisy (i.e. with no modulation embedded) time series data sets obeying to the same characteristics are generated via Monte Carlo, and for each of them the corresponding highest peak in the associated spectrum is recorded. The histogram of these values is used to assess the significance of the highest peak found in the spectrum of the real data: indeed its significance is given by the fraction of times in which a larger value is got in the Monte Carlo calculation, which is thus a measure of the probability to get by pure noise chance a peak as high or higher than that detected experimentally.

Even if not strictly needed, sometimes the Monte Carlo results are fitted to the (50) or (51) in order to get the effective number M of independently scanned frequencies.

Such a Monte Carlo procedure in this paragraph is extensively tested with the purpose to gain adequate insight in it, in view of its subsequent application to the Super-Kamiokande data, also extending the methodology illustrated in the literature to the less high peaks, in order to check how good is the model represented by the (52). To this purpose some example calculations have been performed, referred to an hypothetical series with 100 sampling points, either equally spaced or poisson distributed in time (i.e. with an exponential distribution between two subsequent points).

Just to fix the ideas, the numbers used to describe the fictitious data series in the examples are taken to resemble the Super-Kamiokande data, even if not strictly needed.

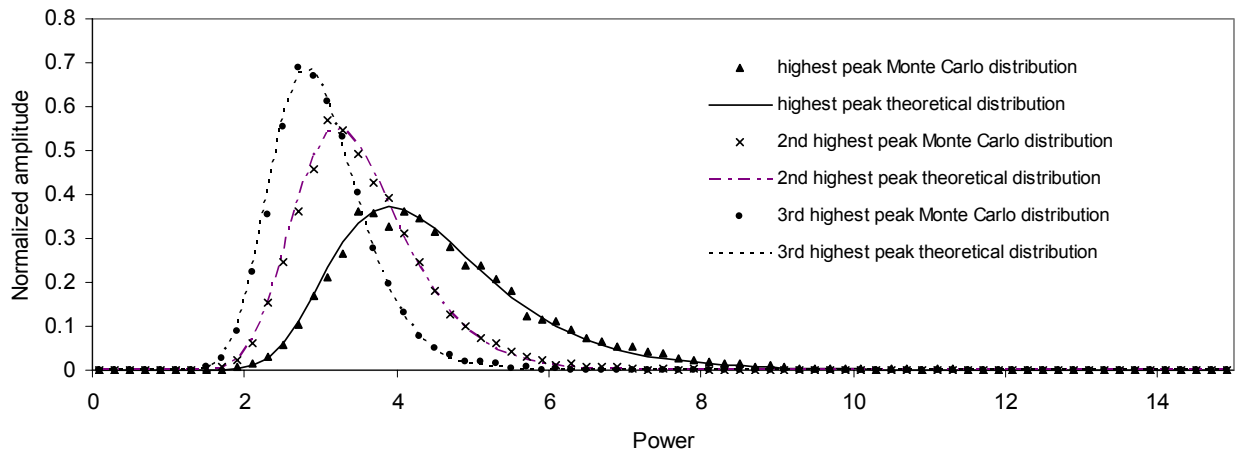

*Fig. 1- Schuster periodogram: comparison of Monte Carlo with the model under the hypothesis of known noise variance*

We remind that the model (52) is expected to be exactly valid in case of even sampling, when the periodogram to be used is the Schuster version, evaluated only at the set of natural frequencies. We can then apply the Monte Carlo procedure to the (42) and check whether the resulting output histograms, describing the distributions of the largest peaks, are in agreement with the (52). The Monte Carlo computation have been accomplished generating for each of the 100 points of the series a random number drawn according to a gaussian distribution of mean value 2.5 and sigma 0.5. For each set of 100 simulated elements, the relevant periodogram has been computed according to the (42), and then the three highest peaks recorded and put in histogram form. The resulting histograms are shown in Fig.1, overlapped to the corresponding model functions: the agreement between the Monte Carlo outputs and the formula (52), computed for M = 50, is really excellent for all the three peaks. Since the (52) are normalized to unit area, for purpose of comparison here and in the following the same normalization is applied to the Monte Carlo distributions.

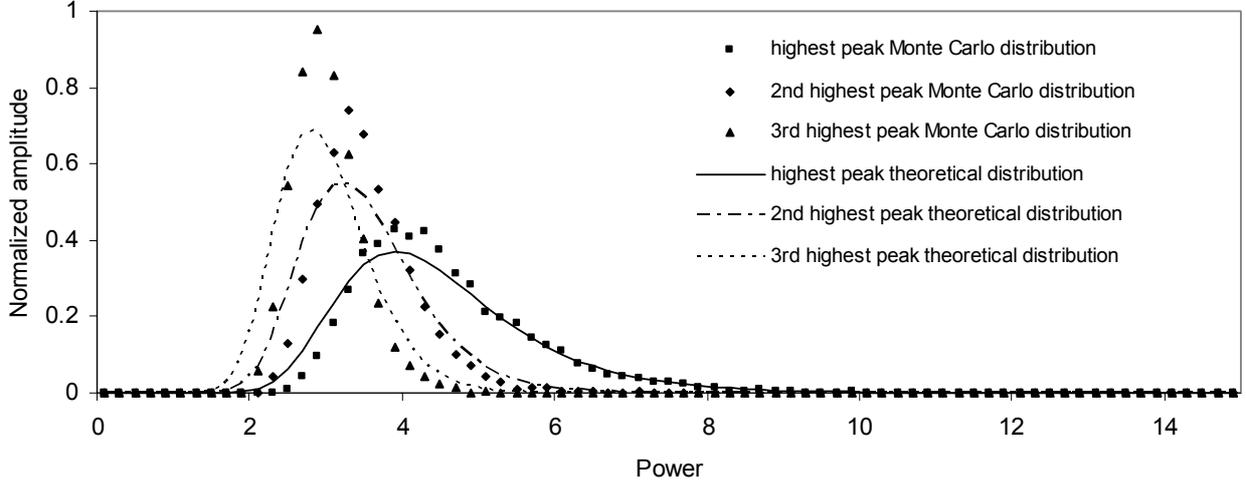

*Fig. 2 - Schuster periodogram: comparison of Monte Carlo with the model under the hypothesis of noise variance estimated from the data*

A caveat, however, is due. The agreement is very good if in the (42) it is inserted as σ the same value used in the generation of the random numbers of the series. Obviously, in a real experimental situation the variance of the data is not known, hence in the (42) it has to be inserted a value estimated from the data itself.

The outcome of such a situation is shown in Fig. 2. The agreement of the Monte Carlo data and of the model is lost, but for the highest peak the tail of the relevant distribution, which is the essential part for its significance assessment, is still properly described by the model function. It must be stressed that the model (52) is independent from the mean value and sigma of the generated numbers forming the noisy series: this fact stems from the normalization effect of the $\sigma^2$ factor at the denominator of the (42) and from the preliminary subtraction of the average value from the terms of the series. The same is true for the Monte Carlo histograms reported in Fig. 2: they depend only on the number of equally spaced data points, as checked changing the values both of the mean value and of the variance of the hypothetical experimental series. It can also be noted that the effect of not knowing a-priori the noise variance is to make the histograms narrower and higher than the corresponding model functions.

As an interesting historical digression, it is worth to note that Fisher in [19] solved exactly the problem of the significance assessment of the highest peak of the Schuster periodogram, in the case of variance inferred from the data itself. Specifically, he proved that, by evaluating the ratio $g$ of the highest periodogram peak to the sum of the ordinates of all the peaks, the PDF of $g$ is given by

$$p(g) = \sum_k \binom{M}{k} (-1)^{k+1} k(M-1)[1-kg]^{(M-2)} \qquad (66)$$

where the sum over $k$ extends up to the minimum between $M$ and the highest integer less than $1/g$. The comparison between the (66) and the corresponding Monte Carlo distribution is reported in Fig. 3: the agreement is excellent.

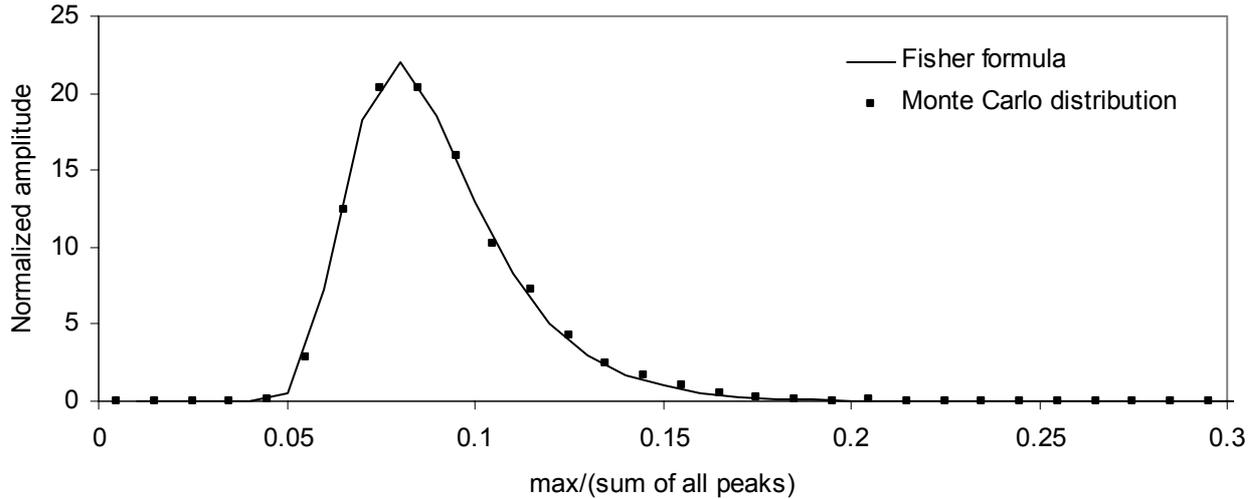

*Fig. 3 - Schuster periodogram: comparison of Monte Carlo with the Fisher formula for the PDF of the ratio g*

From these evaluations we can infer two conclusions: the first is that the Monte Carlo is reliable and well understood in term of analytical models, the second that, since also the lesser amplitude peaks obey to well defined distributions, it is in principle conceivable to attempt a significance assessment also for them, and not only of the highest. This possibility relies on the use either of the model (52) (which practically is only an approximation because of the unknown variance) or, better, of an accurate Monte Carlo calibration like that in Fig. 2. Can a similar approach be pursued also in the case of uneven sampling? To check this possibility, several Monte Carlo computations have been done assuming an hypothetical series of 100 points poisson distributed (with a mean time between two points equal to 5 days).

Such null hypothesis Monte Carlo calculations have been computed adhering to the features of the three different analysis methods under examination: Lomb-Scargle, weighted periodogram and likelihood with asymmetric errors. As for the Schuster periodogram, the starting point of the procedure in all cases is the definition of the characteristics of the hypothetical experimental time series, that provides the input for the subsequent Monte Carlo procedure for significance assessment. Then, in the generation stage, many synthetic time series sets are constructed generating at each point of the series (whose timing retains that of the hypothetical original data series) a random number drawn by a Gaussian distribution with properly chosen mean value and variance.

In particular, in the cases of the Lomb Scargle periodogram the hypothetical experimental series is assumed with mean value evaluated equal to 2.5 and scatter of the measured values equal to 0.5; afterwards in the corresponding Monte Carlo procedure the synthetic time series are generated taking for the Gaussian distributions the same quantities, respectively, for the mean value and the variance. The Lomb-Scargle periodogram for each of these synthetic sets is computed in two different ways: either introducing in the Lomb-Scargle formula as $\sigma$ the generation value 0.5 or the value inferred from the generated series (the latter is obviously the realistic procedure while dealing with real data); as mean value in both cases it is taken the simple average of the synthetic series.

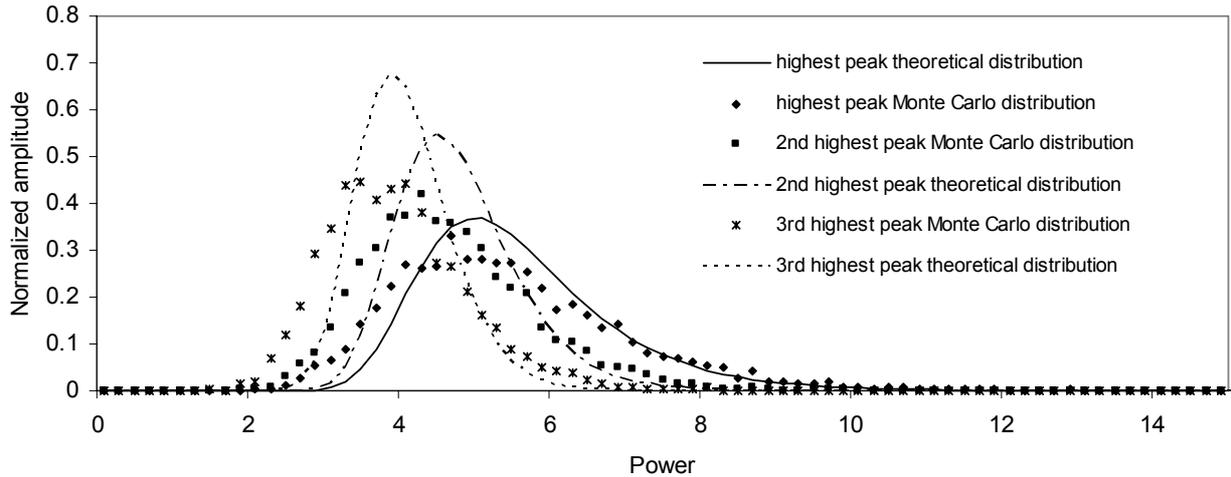

*Fig. 4 - Lomb-Scargle periodogram: comparison of Monte Carlo with the model under the hypothesis of known noise variance*

In the case of the weighted periodogram, the hypothetical experimental series is still assumed with measured mean value 2.5, but to each point it is attributed a fictitious experimental error taken randomly from a distribution centered at 0.5 and with sigma 0.05. In the Monte Carlo procedure the synthetic time series are generated taking for the individual Gaussian distributions the common mean value 2.5 and for the individual variances the corresponding errors at each point. The weighted periodogram for each of these synthetic sets is computed introducing in the weighted periodogram formula the same individual $\sigma_k$'s used in the generation step, while the mean value is taken as the series weighted average.

Finally, in the case of the likelihood analysis done retaining the asymmetric errors, the hypothetical experimental series is still assumed with mean value evaluated to be 2.5, but for each point two fictitious asymmetric errors are assumed (again individually drawn from a distribution centered at 0.5 and with sigma 0.05). In the corresponding Monte Carlo procedure the synthetic time series are generated exploiting asymmetric Gaussian distributions with the upper and lower part parametrized by the corresponding asymmetric errors, and with mean value still equal to 2.5. The likelihood spectrum for each synthetic set is computed with the iterative numerical procedure mentioned in paragraph 3 using the same asymmetric errors exploited in the generation stage, and with mean inferred by the weighted average of the series itself.

In all cases the output of the Monte Carlo are the histograms obtained recording the largest spectral peaks. In the following these histograms for the various cases under investigation are reported , and compared with the model (52).

In Fig 4 and 5 the distributions from the Monte Carlo in the case of the standard Lomb-Scargle periodogram are shown. For coherence with the subsequent calculations of the SK data, the spectral analysis is assumed over the frequency range from 0 to 50 cycles/year. Fig. 4 pertains to the (unrealistic) situation in which the (common) noise variance is perfectly known, while Fig. 5 is related to the more realistic case with the sigma estimated directly from the data, reproducing exactly the standard Lomb-Scargle procedure in a real experimental condition.

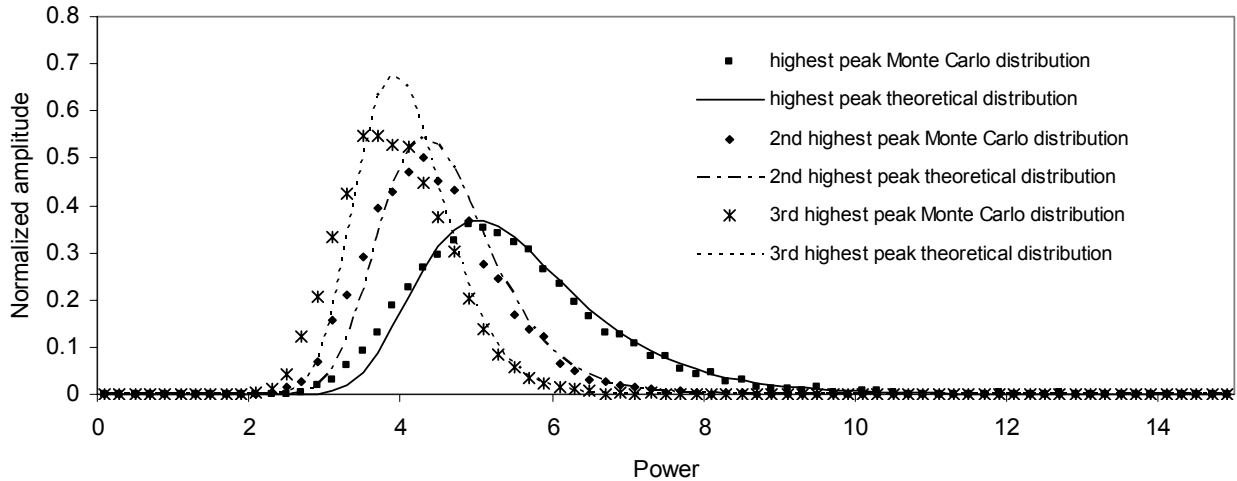

*Fig. 5 - Lomb-Scargle periodogram: comparison of Monte Carlo with the model under the hypothesis of noise variance estimated from the data*

By examining the results it comes out that in the latter case the agreement between the model and the Monte Carlo output is good, and practically perfect for the tails of the distributions: hence the notion of effective independent frequencies is well founded, at least for practical purposes, in the case of the standard Lomb-Scargle methodology. In particular by fitting the highest peak distribution to the model, the value of 137 is inferred for M. For purpose of comparison, also the continuous curves plotted in Fig 4 are referred to the same value of M (a fit is not attempted since the curves deviate significantly from the model).

It can be noted, however, that the tails of the distributions in Fig. 4 are not drastically different from the model functions, especially the tail of the highest peak distribution. It also comes out an interesting difference with the Schuster periodogram: in that case the a-priori knowledge of the variance ensured a perfect match with the model, while now the Monte Carlo histograms evaluated under the same condition are broader and lower than the model functions. On the other hand, the Monte Carlo demonstrates that in the case of the standard Lomb-Scargle approach there is a balance effect between the narrowing of the distributions observed in the case of even sampling, when the variance is not known (Fig. 2), and the broadening of the same distributions in the case of uneven sampling, when the variance is known (Fig. 4). The net effect of this balance is a recovery of a good agreement with the model (52). Interestingly, at least for the distribution of the highest peak the same effect was also noted in [18].

The case of weighted periodogram is shown in Fig. 6.

The results in the figure are similar to those shown in Fig. 4 got using the same single sigma supposed known: the distributions are broader and lower than the model functions, still plotted for M equal 137.

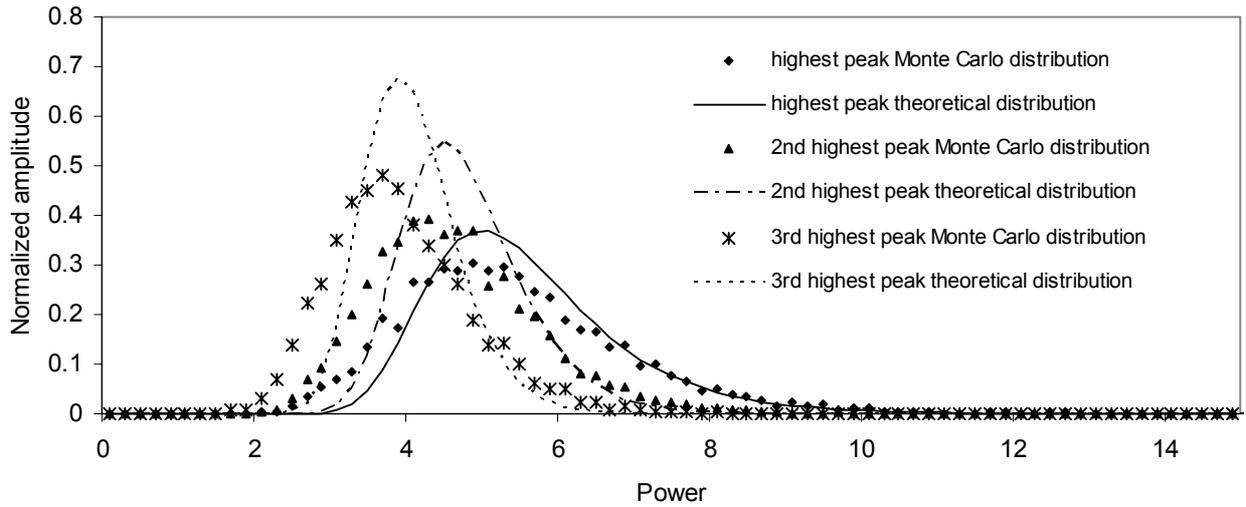

*Fig. 6 – Weighted periodogram: comparison of Monte Carlo with the model. The same variances are used in the generation and in the analysis steps*

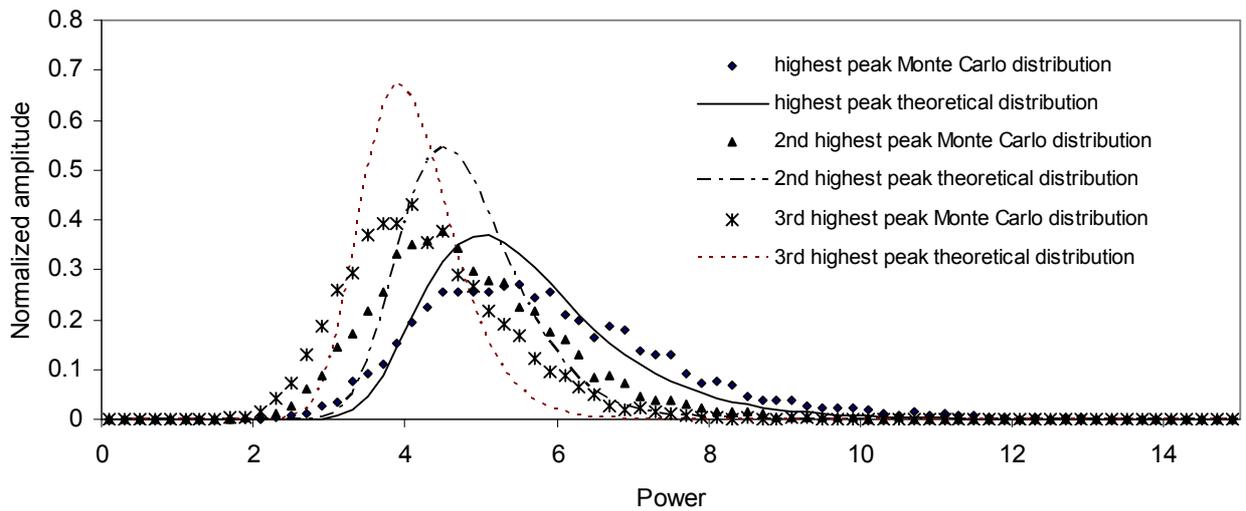

*Fig. 7 – Likelihood spectrum with asymmetric errors: comparison of Monte Carlo with the model. The same asymmetric variances are used in the generation and in the analysis steps*

The histograms in Fig. 7 are those obtained through the application of the likelihood with asymmetric errors; their more remarkable feature is that the respective tails are significantly enhanced with respect to the model functions.

In Fig. 8 the distributions of the highest peak for all the four cases under consideration are displayed together. This way of showing the histograms allows to appreciate clearly that the three situations of variance(s) a priori known are similar (with the asymmetric situation featuring a more pronounced tail), while the Lomb-Scargle standard method histogram profile deviates distinctively from the others; this difference however is less marked in the tail region, where the tail of the Lomb Scargle distribution is only slightly less pronounced than the other tails. Similar conclusions could be drawn by overlapping the corresponding distributions of the lesser amplitude peaks, with the only difference that the tails in the cases of the variances known would appear somehow more enhanced with respect to the Lomb-Scargle histograms.

In summary, the following points can be highlighted as stemming from these Monte Carlo tests: a) also the lesser amplitude peaks and not only the highest obey to well defined distributions;

b) in the standard Lomb Scargle approach such distributions follow well the theoretical model; c) in the other three cases considered the distributions of the various peaks are very similar each other, due to the fact that all such cases share the common feature of supposing the variance (or variances) as known a-priori; d) the tail of the highest peak is similar across the four methods, with the asymmetric case exhibiting the stronger tail; e) the tails of the other peaks are slightly more pronounced in the three cases of variances known with respect to the Lomb Scargle method.

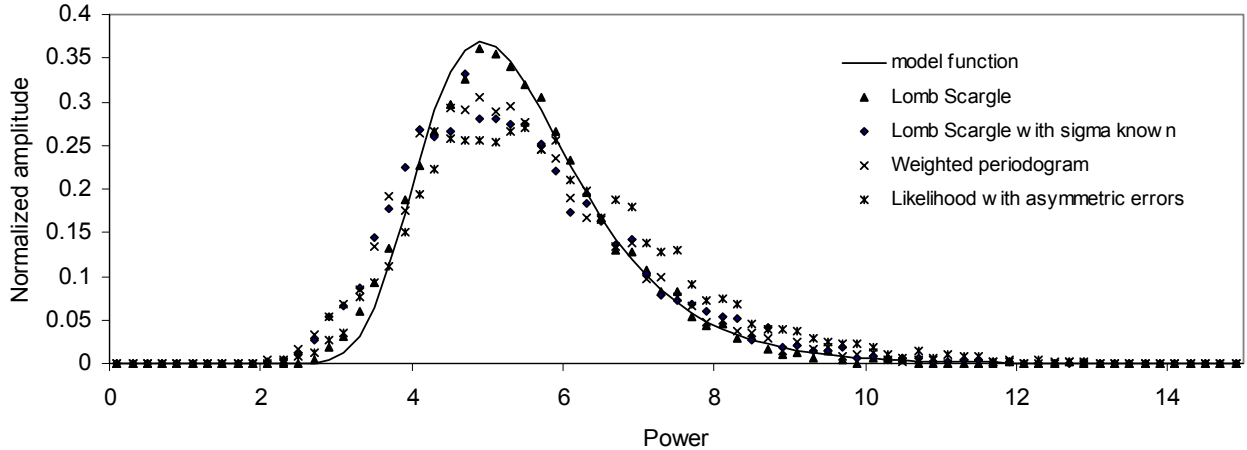

*Fig. 8 – Display of the distributions related to the highest peak in the four calculations described in the paragraph*

The circumstance that the less high peaks feature well defined distributions implies that, not only for the highest peak detected in a experimental spectrum, but also for at least the other more prominent peaks, it can be attempted a significance assessment by comparing their ordinates with the corresponding null hypothesis Monte Carlo distributions. This is not a new idea, since for example it has been studied in [20]. Obviously a multiple peaks significance assessment must be handled with great care: it can be used to reinforce the global agreement, or disagreement, of a spectrum with the constant rate hypothesis, with respect to the information given by the highest peak alone, but it cannot be thought to give automatically indications whether a specific peak is due to signal or noise. Indeed, if there is in a spectrum a mixture of noise and signal lines, what happens is that the correct sequence of the highest, second highest and so on noise peaks is altered by the interleaved presence of the signal lines, thus likely producing one or more inconsistent low values in the assessment procedure, but this inconsistency for example may arise not directly by the signal line or lines, that could be low enough to appear consistent with one or more of the noise distributions, but by a noise peak that, ordered in the wrong way because of the presence of the signal lines themselves, is erroneously confronted with the noise distribution not pertaining to it but to a lower peak. Conversely, if from other information one may peak up correctly the signal lines, then by repeating the significance assessment on the remaining true noise peaks one would recover the correct series of significance values.

Furthermore, the points d) and e) above imply that, given a pure noise series, the spectrum appearance should not change substantially in passing from one of the analysis method to another, especially focusing the attention to the most prominent peaks, and as a consequence the significance values of the highest peaks should be fairly comparable among the different methodologies.

**VIII. DISTRIBUTIONS CORRESPONDING TO A TRUE PERIODICITY**

The previous paragraph has been focused to unravel the features of the null hypothesis distributions. The complementary evaluation is that referred to the alternative situation: given a true periodicity embedded in the time series, what is expected to detect via the spectral analysis? This evaluation can be done both via Monte Carlo or via the numerical integration of the (62).

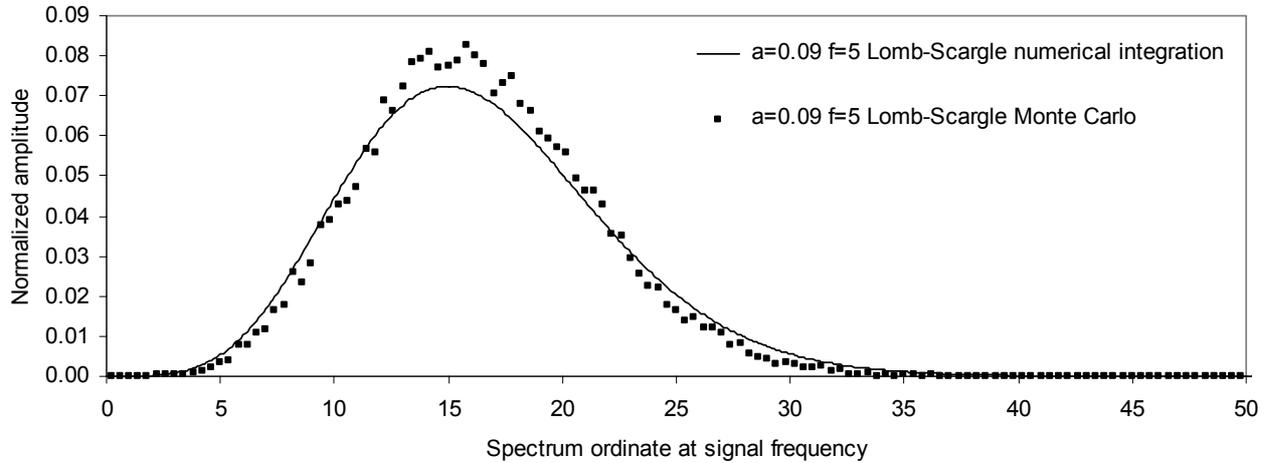

*Fig. 9 – Lomb-Scargle method spectrum ordinate distribution at signal frequency: comparison of Monte Carlo with the model*

The approach through the numerical integration is feasible in the case of the analysis performed either through the Lomb Scargle or the weighted periodogram methods; we remind that in the numerical integration the difference between the two cases is in the definition of the term $X_m$ and $Y_m$, as explained at the end of paragraph 6.

On the other hand, the Monte Carlo procedure is the only one possible in the case of the likelihood approach with asymmetric errors.

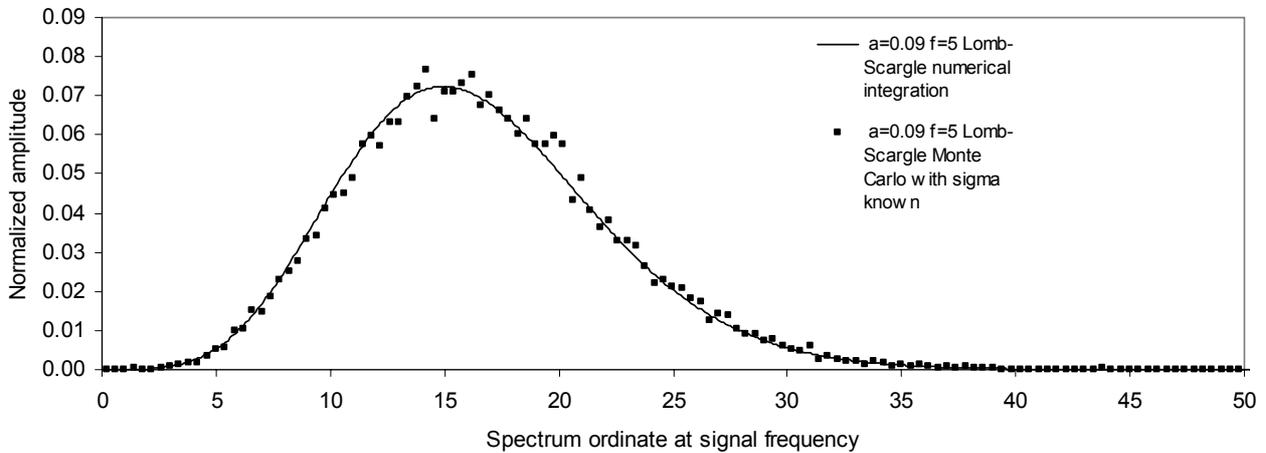

*Fig. 10 – Lomb-Scargle method spectrum ordinate distribution at signal frequency: comparison of the model with the Monte Carlo evaluated assuming sigma known*

To show in a concrete case the behaviour of the spectrum at the frequency of a true signal, and also to compare the Monte Carlo evaluation with the output of the numerical integration, let's assume a frequency equal to 5 cycles/year and an amplitude of oscillation equal to 9% of the average flux and examine the distributions of the relevant spectrum ordinate under different conditions. In order to produce very realistic examples, we consider this periodicity as embedded in a series with timing, mean value and error values equal to that of the 5 day binned SK data set (whose description is postponed to the next paragraph). For the present discussion it must be kept in mind only that each point of the series is given with two experimental asymmetric errors.

In figure 9 there is the plot of the height distribution of the spectral ordinate at the signal frequency, obtained through the numerical integration of the (62) in the case of the Lomb Scargle periodogram, overlapped to the corresponding Monte Carlo distribution, computed generating the time series via the experimental asymmetric errors and performing the analysis on each generated data set through the Lomb Scargle procedure. The Monte Carlo follows well the curve, but not perfectly, the reason being equal to that already encountered at the beginning of the discussion of the null hypothesis distribution for the Schuster periodogram: the normalization sigma at the denominator of the Lomb Scargle formula is itself a random term, since it is inferred from the data and it is not known a-priori, thus adding a variability effect which is not contemplated by the (62). To check that this is actually the origin of the discrepancy, it has been done a test changing the generation procedure in the Monte Carlo by using a single standard deviation for all the points, and then using exactly the same value in the subsequent Lomb Scargle stage as normalization sigma; in this way the Monte Carlo and the numerical integration plot should coincide perfectly. Such a perfect coincidence is indeed shown in Fig. 10, which hence provides the confirmation that both the Monte Carlo and the model are reliable and well understood.

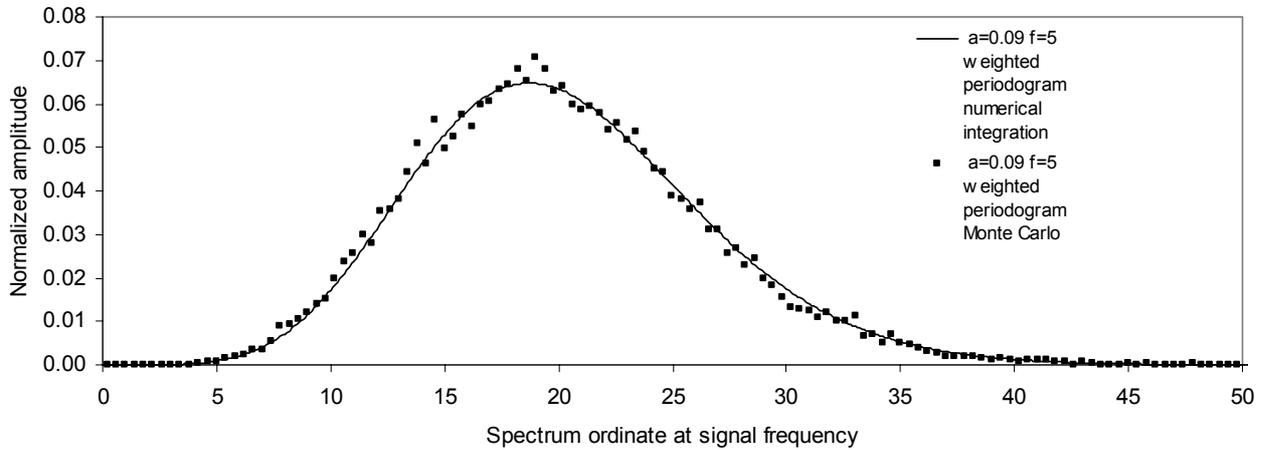

*Fig. 11 – Weighted periodogram method spectrum ordinate distribution at signal frequency: comparison of the model with the Monte Carlo*

In Fig 11 the curves related to the weighted periodogram case are reported: in the Monte Carlo the time series is generated exploiting, as before, the asymmetric errors, while in the computation of the weighted periodogram for each data set the variance at every point is taken as the mean of the two experimental asymmetric errors. The agreement in this case between the Monte Carlo and the numerical model is very good.

From the plots in Fig. 9 and 11 it can be inferred that the error done replacing the Monte Carlo with the numerical integration of the (62) is minimum, with the benefit on the other hand of speeding up the computations. Furthermore, confronting directly the Lomb Scargle and weighted periodogram cases for the same frequency and amplitude, as in Fig 12, it stems the important result that the distribution of the ordinate of the spectrum corresponding to a true periodicity shifts toward higher values in passing from the former to the latter method.

To complete this review it is left to show the distribution obtained with the third method contemplating the analysis via the likelihood with asymmetric errors. Such a Monte Carlo distribution is reported in Fig 13, together with the numerical computation related to the weighted periodogram case.

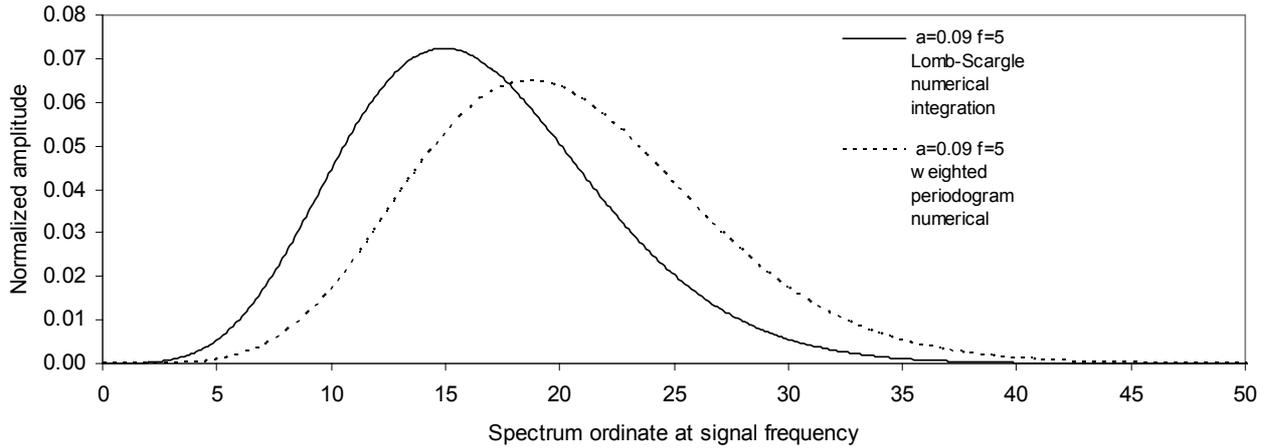

*Fig. 12 – Comparison of the spectrum ordinate distributions at signal frequency related to the Lomb Scargle and weighted periodogram methods*

It can be seen that the Monte Carlo predicts that the inclusion of the asymmetric errors produces a modification of the distribution of the spectrum ordinate at signal frequency with respect to that of the weighted periodogram case, in which the errors are included after being made symmetric via their average. In particular the net effect is a certain shift of the whole distribution toward higher values.

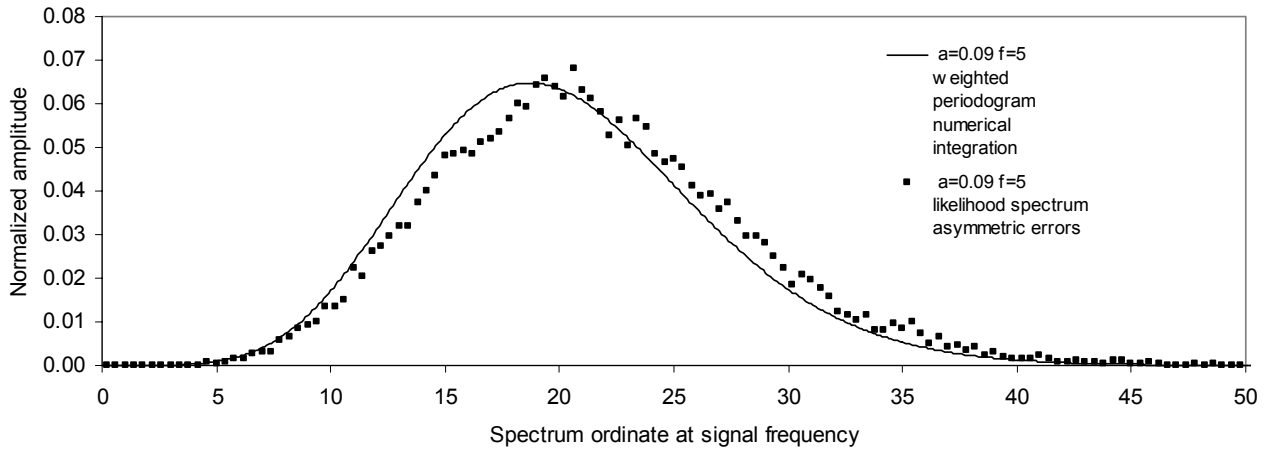

*Fig. 13 – Comparison of the spectrum ordinate distributions at signal frequency related to the weighted periodogram and likelihood with asymmetric errors methods*

Equipped with this model and Monte Carlo insight into the spectrum distribution properties both in case of the null hypothesis and in presence of a true periodicity, the rest of the paper is devoted to analyze in detail the actual SK data.

### IX. ANALYSIS OF THE SK 10 DAY BINNED DATASET: ASSESSMENT OF PEAKS SIGNIFICANCE VIA COMPARISON WITH THE MONTECARLO NULL HYPOTHESIS DISTRIBUTIONS

The Super-Kamiokande collaboration published the time series of the $^8$B neutrino flux measurements organized both in 10 and 5 day bins [1]. For each bin it was provided, together with the flux value, also the respective mean live time $t_{wk}$, properly evaluated in order to take into account the live time of the detector, as well as the relevant corrective factor to remove the 7% peak-to-peak annual variation due to the Earth's orbit eccentricity around the Sun. However, in all the calculations reported in the following it has been preferred not to apply such a correction, in order to leave the data fully intact. The Super-Kamiokande collaboration published also for each bin the two asymmetric errors $\sigma_k^{up}$ and $\sigma_k^{down}$; as thoroughly explained above, in the standard Lomb-Scargle periodogram they are ignored and replaced with a common $\sigma$ given by the scatter of the measured values. On the contrary, in the weighted periodogram approach the symmetric $\sigma_k$'s are taken as

$$\sigma_k = \frac{\sigma_k^{up} + \sigma_k^{down}}{2}$$

while in the likelihood method with asymmetric errors the $\sigma_k^{up}$ and $\sigma_k^{down}$ are taken as they are.

In addition, the datasets provide for each 10 or 5 day long segment the start time $t_s$ and the end time $t_e$.

In this paragraph it is reported for the 10 day binned dataset the null hypothesis analysis in the framework of the three type of investigations under consideration. In particular, for each case it is shown the respective power spectrum as well as the null hypothesis Monte Carlo distributions for the largest spectral peaks, from which the significances of the actual highest peaks in the spectrum are derived.

Adhering to the guidelines of the previous paragraph 7, the Monte Carlo distributions are evaluated differently according to the features of the three different analysis methods. So, specifically, in all cases the synthetic time series sets are constructed generating at each point of the series (whose timing retains that of the original data series) a random number drawn by a Gaussian distribution of specific mean value and variance: in the case of the Lomb Scargle periodogram, the mean value is equal to the average value of the experimental series, and the variance is common for all the points and equal to the scatter of the values of the experimental series; in the case of the weighted periodogram, the mean value is taken equal to the weighted average of the data series, while the variances are different point by point and taken equal to the mean value of the two corresponding asymmetric errors; finally, in the case of the analysis done retaining the asymmetric errors, the individual gaussian distributions used to generate the Monte Carlo series are kept asymmetric, with the upper and lower part parametrized by the corresponding asymmetric errors. After the generation process, the Monte Carlo proceeds as described in the paragraph 7, producing as output the desired null hypothesis distributions of the largest spectral peaks. The results are thoroughly described in the following.

### A. 10 Day Data Lomb-Scargle Periodogram

The Lomb-Scargle periodogram of the 10 day binned SK data is shown in Fig. 14, in the frequency range from 0 to 50 cycles/year.

The four highest peaks have ordinates respectively 7.1 (26.51), 6.8 (26.99), 6.41 (9.4) and 5.36 (23.6) (in the brackets there is the indication of the respective frequency expressed in cycles/year; in the rest of the paper the frequencies are normally written leaving implicit the unit of cycles/year). Here and in the following, to each line should be considered attached an uncertainty of ± 0.07, corresponding to the FWHM of the spectral lines. From a Monte Carlo similar to that used to derive the Fig. 5 in the paragraph of the examples, the results reported in Fig. 15 are obtained. It can be seen that the distribution of the highest peak is very well described by the model function plotted with M =529.

The distributions of the other three less high peaks are somehow also described by the same M value, even if the tails are not perfectly matched. One may wonder why the agreement between the model and the Monte Carlo is better here than in the example reported in Fig. 5. The reason is likely in the different spacing of the points of the series: while in the example in Fig. 5 the points were poisson distributed in time, in the case of the SK data the spacing between them maintains a high degree of regularity, even though is not perfectly even.

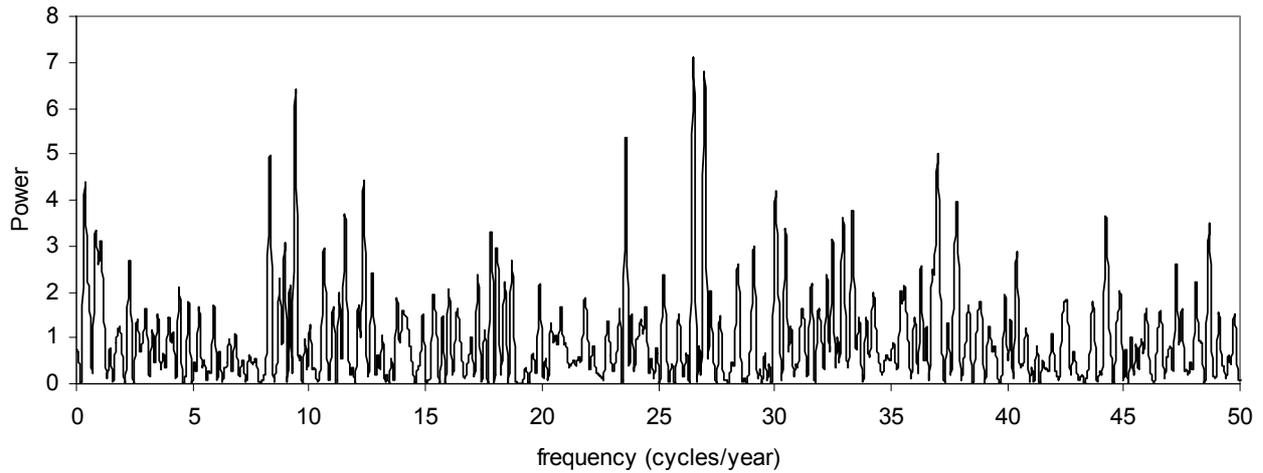

*Fig. 14 – Lomb Scargle periodogram of the 10 day binned SK data*

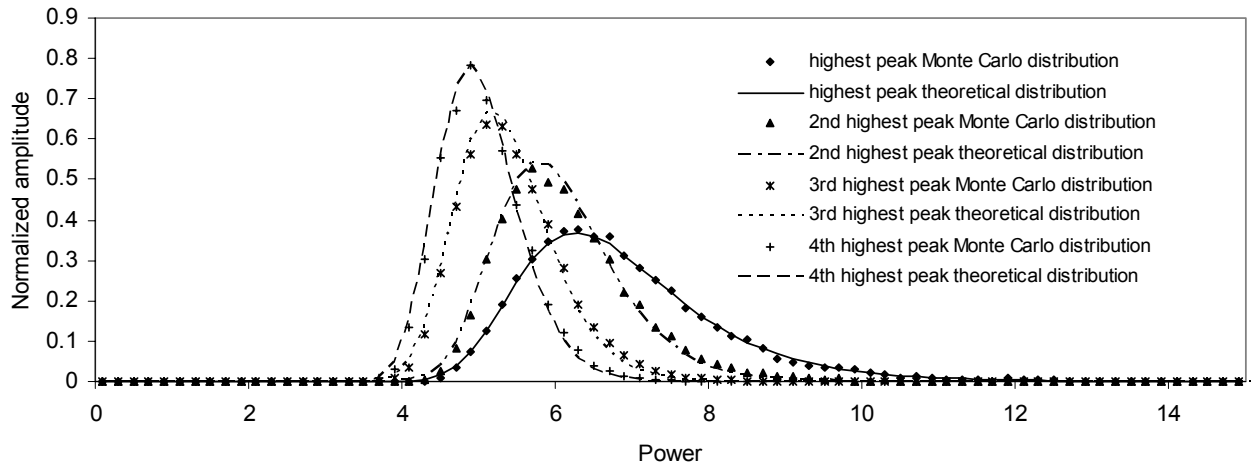

*Fig. 15 – Lomb-Scargle periodogram of the 10 day binned SK data: comparison of the null hypothesis Monte Carlo distributions with the model, noise variance estimated from the data*

The four major peaks listed above are well consistent with the simulated distributions, in the sense that none of them lies on the tail of the respective distribution. Quantitatively, reminding that the definition of significance of a peak is the integral of the corresponding Monte Carlo distribution above its actual value, we get that for the peak at 7.1 the significance is 34.7 % (3474 entries above the 7.1 ordinate out of 10000 simulated events), that for the peak at 6.8 the significance is 15% (1504 entries above the 6.8 ordinate out of 10000 simulated events), that for the peak at 6.41 the significance is 8% (802 entries above the 6.41 ordinate out of 10000 simulated events), and that for the peak at 5.36 the significance is 27% (2694 entries above the 5.36 ordinate out of 10000 simulated events). So, in summary, the Lomb Scargle periodogram of the 10 days binned data is perfectly consistent with a noisy series with no periodicity embedded. This result is in agreement with the conclusions in [1] got through the same analysis method, limited to consider the highest peak only.

**B. 10 Day Data Weighted Periodogram**

The weighted periodogram of the 10 day binned SK data is shown in Fig. 16, in the frequency range from 0 to 50 cycles/year. The four highest peaks have ordinates respectively 8.5 (26.52), 7.46 (9.4) 6.84 (26.98) and 6.01 (23.61). Hence these are the same four peaks identified in the Lomb-Scargle spectrum, with the difference that the second highest peak here was the third highest in the previous spectrum and viceversa.

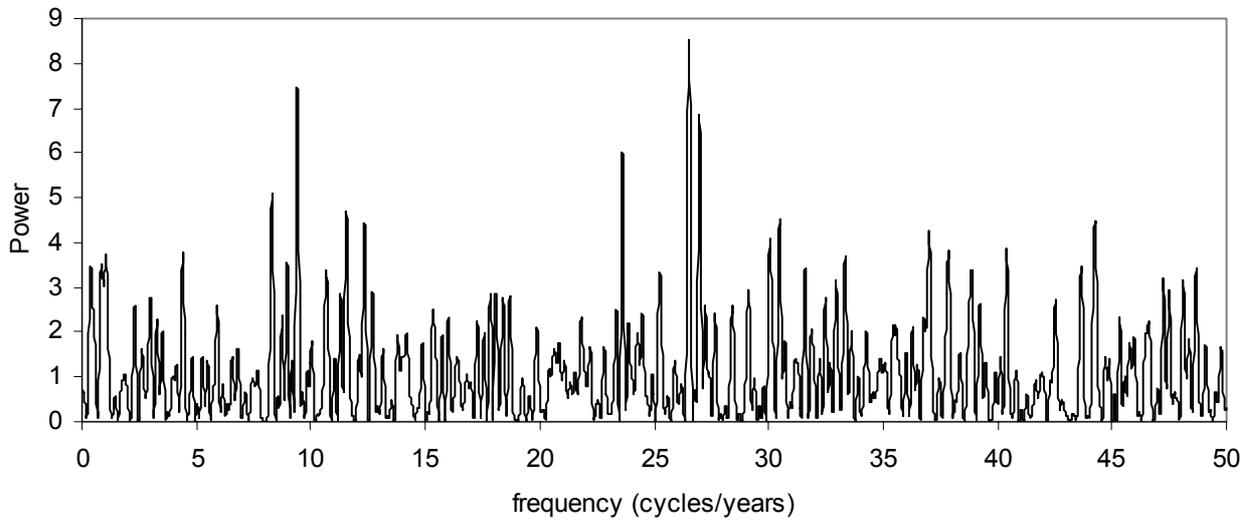

*Fig. 16 – Weighted periodogram of the 10 day binned SK data*

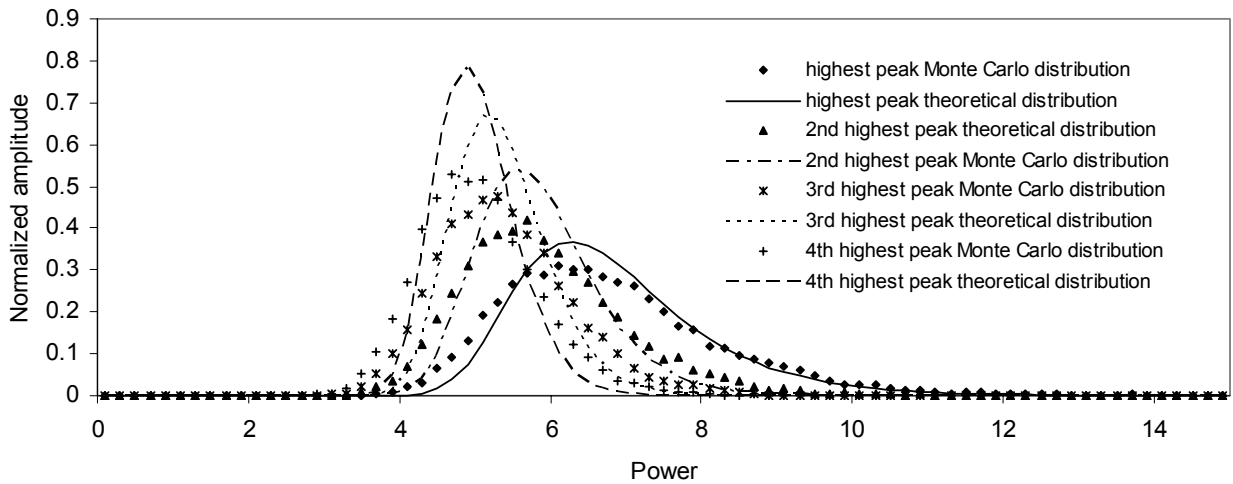

*Fig. 17 – Weighted periodogram of the 10 day binned SK data: comparison of null hypothesis Monte Carlo distributions with the model*

From a Monte Carlo similar to that used to derive the Fig. 6 we get the results reported in Fig. 17. It can be seen, as in the example of Fig. 6, that the Monte Carlo histograms do not follow the model functions, reported in the figure for the same value M=529 derived above. Quantitatively, from the Monte Carlo distributions we get that for the peak with ordinate 8.5 the significance is 12.4 % (1239 entries above 8.5 out of 10000 simulated events), that for the peak with ordinate 7.46 the significance is 9.1% (912 entries above the 7.46 ordinate out of 10000 simulated events), that for the peak with ordinate 6.84 the significance is 6.5% (654 entries above the 6.84 ordinate out of 10000 simulated events), and that for the peak with ordinate 6.01 the significance is 11.2% (1125 entries above the 6.01 ordinate out of 10000 simulated events).

So, also the weighted periodogram of the 10 days binned data appears to be consistent with a noisy series with no periodicity embedded. There are however some facts that deserve to be pointed out: the first is the inversion of the order of the second and third peak, the second is a general increase trend of the ordinate of the four highest peaks which finds its counterpart in the overall decrease of the significance values with respect to the previous periodogram, the third is the enhancement of the tails of the null hypothesis simulated distributions (actually such an enhancement is more marked for the lowest peaks, while the tail of the highest is pretty unchanged). A thorough discussion, encompassing also these points, is deferred to the next paragraph 11.

### C. 10 Day Data Likelihood Spectrum With Asymmetric Errors

The likelihood spectrum with asymmetric errors of the 10 day binned SK data is shown in Fig. 18, in the frequency range from 0 to 50 cycles/year. The four highest peaks have ordinates respectively 8.08 (26.51), 7.05 (9.4), 6.44 (26.98) and 5.67 (23.61), thus the same peaks as before.

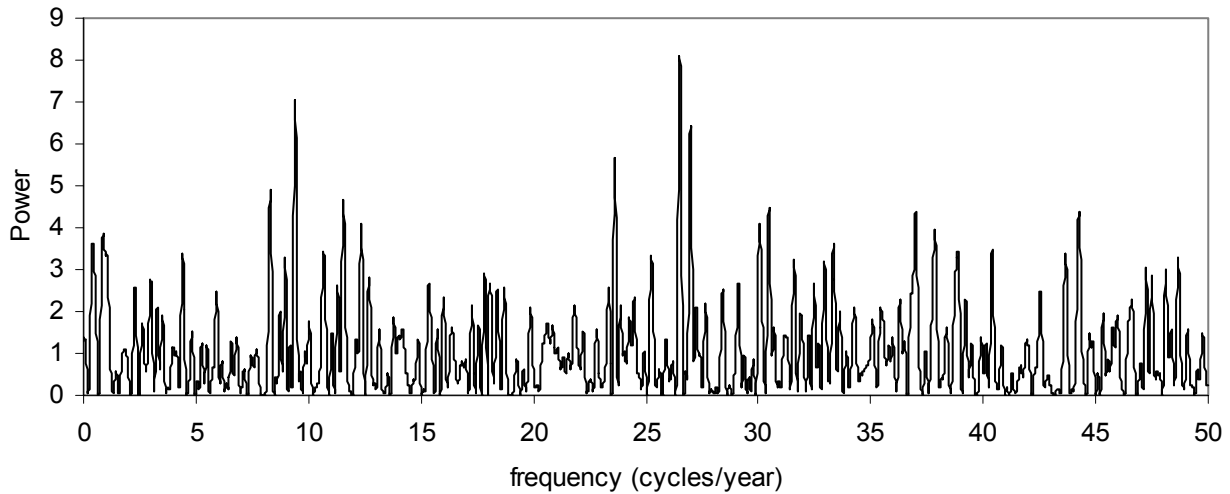

*Fig. 18 – Likelihood spectrum with asymmetric errors of the 10 day binned SK data*

From a Monte Carlo similar to that used to derive the Fig. 7 in the example paragraph we get the results reported in Fig. 19. It can be seen, as in the example of Fig. 7, that the Monte Carlo histograms do not follow the model functions, plotted for comparison in the figure with M=529.

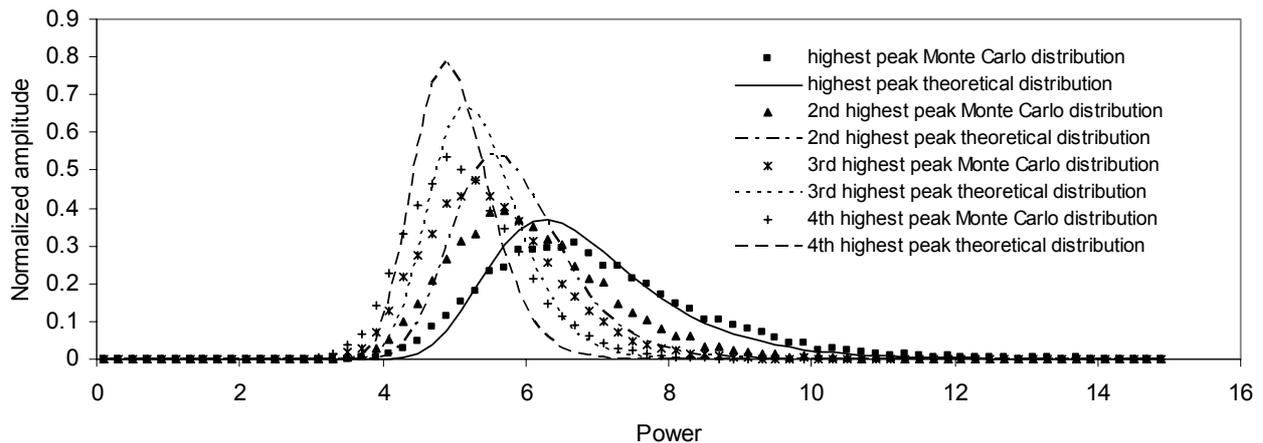

*Fig. 19 – Likelihood spectrum of the 10 day binned SK data: comparison of null hypothesis Monte Carlo distributions with the model*

Quantitatively, from the Monte Carlo histograms we get that for the peak with ordinate 8.08 the significance is 20.1 % (2013 entries above 8.08 out of 10000 simulated events), that for the peak with ordinate 7.05 the significance is 18.2% (1824 entries above the 7.05 ordinate out of 10000 simulated events), that for the peak with ordinate 6.44 the significance is 16.1% (1611 entries above the 6.44 ordinate out of 10000 simulated events) and that for the peak with ordinate 5.67 the significance is 25.6% (2559 entries above the 5.67 ordinate out of 10000 simulated events). So, what happens is that the ordinates of the highest peaks are larger than in the Lomb-Scargle case, but lower than in the weighted periodogram spectrum; moreover, the tails of the null hypothesis distributions are enhanced with respect to both the previous case, and as a consequences the resulting significances of the highest peaks are well consistent with a noisy series with no periodicity embedded. It should be, however, pointed out that this last spectrum as appearance is more similar to the weighted periodogram spectrum than to the Lomb-Scargle one, in particular for what concerns the sequence of the highest peaks.

Prior to further discuss the features of the three spectra under consideration in paragraph 11, in the next paragraph we exploit the calculations of paragraph 8 to check what would be the predictions for a true periodicity embedded in the data.

## X. ANALYSIS OF THE SK 10 DAY BINNED DATASET: PREDICTION OF THE SPECTRUM BEHAVIOUR IN PRESENCE OF A TRUE PERIODICITY

By referring to the calculations reported in paragraph 8 it is known that in case of a true periodicity embedded in the data what is expected is, at the signal frequency, a spectral ordinate which obeys to the distribution (62) (see examples in paragraph 8). In order to summarize meaningfully in the case of the 10 day binned data the expected spectral response to a true periodicity over the search band, in Fig. 20 it is plotted the mean value of the signal induced distribution (62) as function of the frequency.

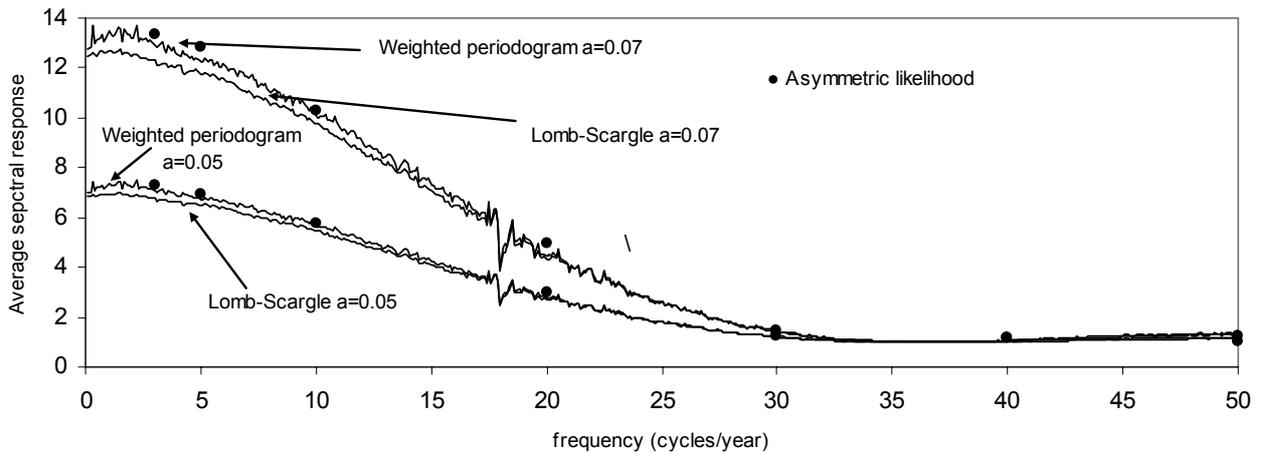

*Fig. 20 – Average spectral response over the whole search band*

The curves are computed for the two different relative amplitudes of 0.05 and 0.07, and for the cases of Lomb-Scargle and weighted periodogram. Few dots are added to show also the behaviour of the likelihood spectrum with asymmetric errors. Some things deserve to be highlighted: a) the spectral response is not uniform over the search band and vanishes rapidly beyond the Nyquist frequency (signalled by the irregularity present on all the curves); b) the spectral response changes sharply as function of the amplitude of the periodicity; c) in passing from the first to the second and third method, in average the spectrum ordinate increases slightly.

Another prediction is that related to the possible alias. It has been suggested in previous analysis [2][3] that the two lines at 9.4 and 26.52 can be an alias pair. Through the (62) this can be checked quantitatively: assuming a periodicity of a given frequency and amplitude the (62) is computed for all the frequencies of the search band, in order to determine at frequencies different from the true frequency what is the respective amplitude distribution of the spectrum ordinate induced by that true frequency. At all frequencies, with exception of the alias ones, such a computation should simply produce the expected noise exponential distribution, while at the alias frequency (or frequencies) the same calculation should reproduce a distribution similar to that induced by a real signal.

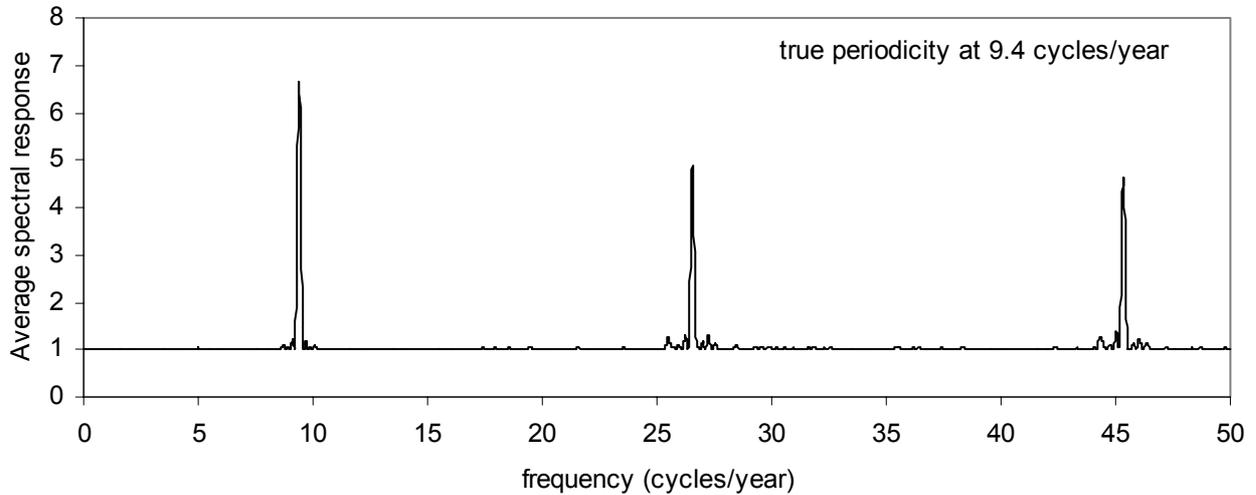

*Fig. 21 – Predicted alias frequencies of a real modulation with frequency equal to 9.4 cycles/year*

In order to understand which is the case, for every computed distribution it is evaluated the respective mean value, that must be 1 for the non alias frequencies (mean value of the expected exponential distribution), and higher than 1 for the alias frequencies. A plot that reports the mean value against the frequency will thus display sharp peaks at the alias frequencies. The result of this search is plotted in Fig. 21, for a presumed periodicity of 9.4 cycles/year with relative amplitude 0.055. Two alias frequencies are predicted, one at 26.55 and the other at 45.34.

In the next paragraph these results and those of paragraph 9 are examined together in the attempt to construct a consistent, explicative picture.

## XI. OVERALL COMPARISON OF THE MODEL PREDICTIONS WITH THE DATA

Having studied in paragraph 9 the expected properties of the null hypothesis distributions and in paragraph 10 the properties of the spectral responses in case of a true periodicity, it can now be tried to outline an overall interpretation of the detected behaviour of the 10 day SK spectra obtained with the various methods of analysis.

Prior to go ahead with such a discussion it is worth to note that, while doing the analysis thoroughly explained above, the fit gives an estimate of the relative amplitude associate to each spectral line. In particular for the line at 9.4 it results: in the Lomb Scargle analysis an amplitude of $0.052 \pm 0.014$, in the weighted periodogram analysis an amplitude of $0.054 \pm^{0.015}_{0.014}$ and in the likelihood with asymmetric errors an amplitude of $0.054 \pm^{0.015}_{0.014}$. So, we are dealing with a potential effect in the 5-6% amplitude range.

Coming back to the overall interpretation attempt, the peculiar fact detected experimentally while computing the spectra under the three different methods is the increase of the ordinates of the more prominent peaks while passing from the standard Lomb-Scargle methodology to the two methods taking into account the errors. However, the increase is more evident in the weighted periodogram case than in the asymmetric errors spectrum. On the other hand, the Monte Carlo evaluation shows that the tails of the null hypothesis distributions, when compared with the Lomb-Scargle configuration, change slightly in the weighted periodogram case while are markedly enhanced in the asymmetric likelihood configuration. Thus it stems that the significance assessment of the peaks becomes stronger in the second method with respect to the standard Lomb-Scargle methodology, while it is pretty comparable with it in the third approach.

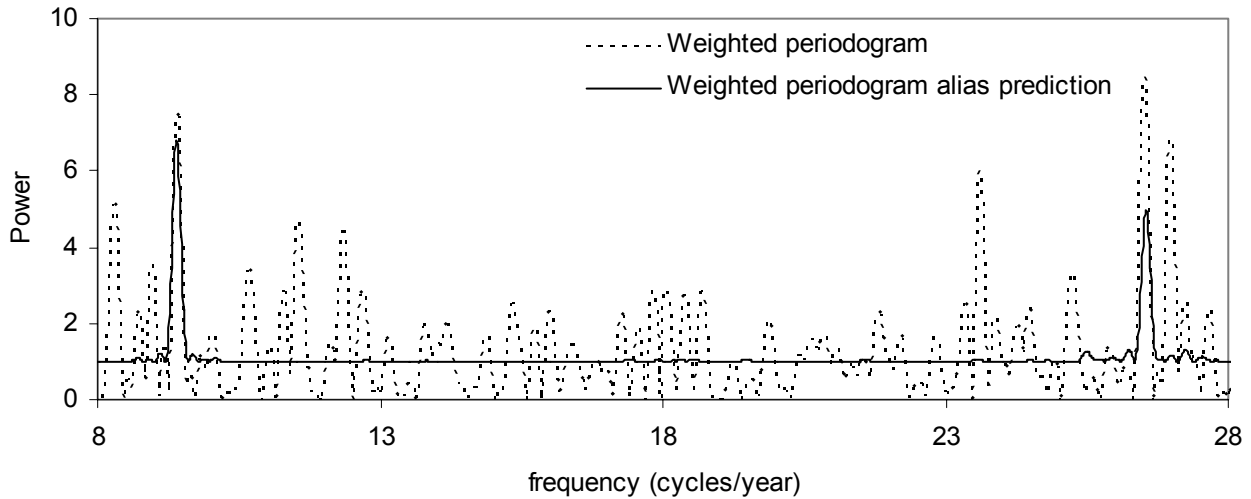

*Fig. 22 – Plot of the alias prediction for a true periodicity at 9.4 cycles/year overlapped on the actual spectrum (weighted periodogram case)*

Very interesting is the comparison of the spectra with the predicted spectral response in Fig. 20. Three of the four more prominent peaks, those at frequency 23.61, 26.52 and 26.98, fall in a region where the spectral analysis has reduced sensitivity to a true periodicity; this quantitative observation induces to presume that they may arise from fake effects.

In particular, the above alias calculation suggests that the peak at 26.52 could well be the alias of a real signal located at 9.4 cycles/year: the alias predicted by the calculation is indeed at 26.55, so practically coincident with the location of the detected peak at 26.52 (see Fig. 22). The alias calculation, however, predicts also another peak at 45.34. In order to locate with respect to the Nyquist frequency such a peak, let's consider that the first alias frequency must be specularly located with respect to the true signal around the Nyquist frequency. Hence, denoting such a frequency with Nq, we have that (Nq-9.4)+Nq=26.55 and hence Nq=(26.55+9.4)/2=17.975, very close to the nominal value of 18. From this result it stems that 2x17.975+9.4 is equal to 45.35, coincident with the second alias predicted by the calculation, which is therefore an alias shifted on the right of the double of the Nyquist frequency of a quantity just equal to the 9.4 frequency.

The investigation of the spectra shows that actually also this second peak is present, with low height in the Lomb-Scargle case and slightly more pronounced in the other two spectra.
Moreover, the strong correlation of the two peaks at 9.4 and 26.52 emerges also from another test, whose output is shown in Fig. 23: the line at 9.4 is removed from the times series, exploiting its fit parameters (hence not only the amplitude but also the phase), and then the spectrum re-computed. From Fig. 23, in which the original spectrum is reported as dashed line, and the spectrum after the subtraction as solid line, it is inferred that the subtraction of the 9.4 line automatically removes also the 26.52 line.

It must be, however, pointed out that these plurality of alias indications are necessary conditions that must be met if it is a priori known that there is a real modulation, but that viceversa, taken alone, cannot be considered as a proof for the presence of a signal. The reason is that among the noise peaks it exists a correlation that can mimics the alias effect itself. To show such a correlation it has been performed a Monte Carlo test generating a sets of purely noise series, identifying the maximum of the spectrum within the first Nyquist interval, and then recording the height of the spectrum at the alias frequency. The result of such a test is reported in Fig. 24, where the plots displayed are respectively the histogram of the maximum within the first Nyquist interval and the histogram of the corresponding spectrum ordinate at the alias frequency; clearly the latter histogram is very different from the usual simple exponential distribution that would have emerged in case of absence of correlation and shows that the probability of higher values is much enhanced with respect to the normal noise distribution.

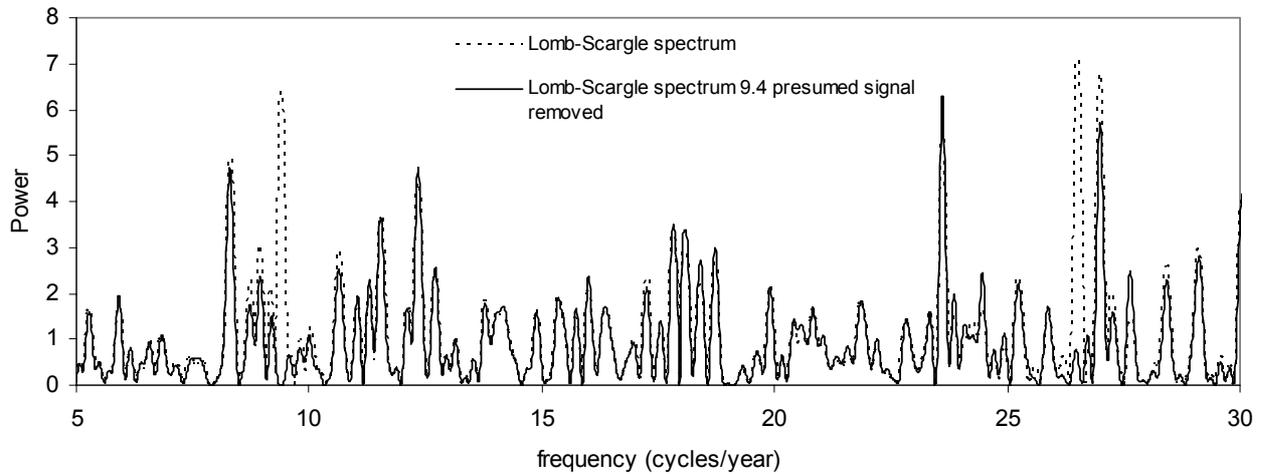

*Fig. 23 – Lomb-Scargle spectrum recomputed (solid line) after subtraction of the presumed 9.4 line*

Thus the possible alias relation between the 26.52 and the 9.4 line could be considered only a working hypothesis, to be confirmed, or disproved, in a more general framework. In particular in this respect it will be important to check the behaviour of the 5 days dataset (see next paragraphs). While waiting for that test, what can be done here is to hypothesize that three frequencies 23.61, 26.52 and 26.98 are truly artefacts, with the 26.52 being the alias of that at 9.4, and then, following the suggestion in the paragraph 7, repeat the significance assessment calculation ignoring the peaks suspected to be induced by real signals, so to check if the residual peaks produce an overall global significance more constant across the three spectra.

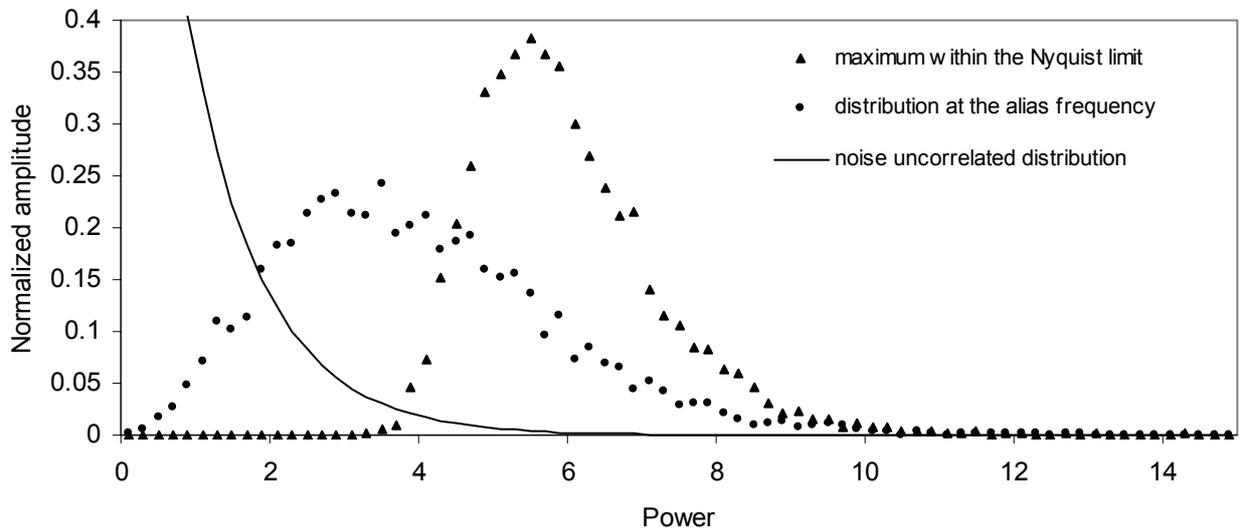

*Fig. 24 – Correlation of the alias pairs in the spectrum of a purely noisy series*

Clearly, in such a scenario it is presumed that the 9.4 line is a true signal. What is, however, not consistent with the model is the circumstance that for a true periodicity the calculation predicts, in average, a modest increase of the spectrum ordinate while passing from the first to the second and third method, while in the real spectra the increase of the 9.4 ordinate is definitely more pronounced. Interestingly, in this respect the disagreement with the model is less pronounced in the asymmetric likelihood case.

Keeping in mind this caveat, we can in any case do the significance exercise; the significances of four highest peaks, ignoring the 9.4 line and its presumed alias, are summarized in the following tables 1, 2 and 3.

| Ordinate | Frequency | Significance |
| --- | --- | --- |
| 6.80 | 26.99 | 4393 events out of 10000 |
| 5.36 | 23.60 | 7265 events out of 10000 |
| 5.00 | 37.00 | 7158 events out of 10000 |
| 4.98 | 8.31 | 5187 events out of 10000 |

*Tab 1 – Lomb Scargle periodogram: peak significance ignoring the presumed signal at 9.4 cycles/year and its alias*

| Ordinate | Frequency | Significance |
| --- | --- | --- |
| 6.84 | 26.98 | 4264 events out of 10000 |
| 6.01 | 23.61 | 4103 events out of 10000 |
| 5.10 | 8.30 | 5995 events out of 10000 |
| 4.7 | 11.55 | 6468 events out of 10000 |

*Tab 2 – Weighted periodogram: peak significance ignoring the presumed signal at 9.4 cycles/year and its alias*

| Ordinate | Frequency | Significance |
| --- | --- | --- |
| 6.44 | 26.98 | 5942 events out of 10000 |
| 5.67 | 23.61 | 6021 events out of 10000 |
| 4.89 | 8.30 | 7504 events out of 10000 |
| 4.67 | 11.55 | 7252 events out of 10000 |

*Tab 3 – Asymmetric likelihood: peak significance ignoring the presumed signal at 9.4 cycles/year and its alias*

Therefore in all three cases the four residual highest peaks would produce a good overall agreement with the hypothesis of being consistent with a noisy spectrum, fairly comparable among the three spectra, in contrast with the not uniform behaviour throughout the same spectra (particularly in the second case) detected before while considering also the 9.4 line and its potential alias. As a consistency check, the same exercise has been repeated after subtracting from the time series data the presumed modulation at 9.4 cycles/year, getting comparable results. From the examination of the tables it stems that also the absolute value of the residual highest peaks are very uniform throughout the three spectra, in contrast with the observed behaviour of the 9.4 line and of its potential alias.

Hence from the summary of the evaluations presented here, based essentially on the indications of the alias calculation and of the comparison with the model predicting the increase of the ordinate of the line corresponding to a true signal in the methods taking into account the errors, the suggestive possibility emerges of a true periodicity at 9.4 accompanied by an alias at 26.52, with the other peaks being "well behaving" noise peaks. Such a scenario, however cannot be pushed beyond a simple working hypothesis, because the significance calculation does not support this possibility, being in any case all the three original spectra well consistent with the hypothesis of constant rate. Furthermore, it must be stressed that the actual increase of the ordinate of the 9.4 line in passing from the first method to the second and third ones, is larger than the average increase predicted by the model in the same transitions for a true signal line, which is evaluated to be about 5% in both cases, whilst in the data it amounts respectively to 16% and 10%.

## XII. ANALYSIS OF THE SK 5 DAY BINNED DATASET: ASSESSMENT OF PEAKS SIGNIFICANCE VIA COMPARISON WITH THE MONTE CARLO NULL HYPOTHESIS DISTRIBUTIONS

The characteristics of the 5 day binned dataset released by the Super-Kamiokande collaboration are similar to those of the 10 day case and therefore are not repeated here. In this paragraph we present, in total analogy with the 10 day case, the features of the null hypothesis distributions and their comparison with the more prominent peaks in the respective experimental spectra.

### A. 5 Day Data Lomb-Scargle periodogram

The Lomb-Scargle periodogram of the 5 day binned SK data is shown in Fig. 25, in the frequency range from 0 to 50 cycles/year.

The three highest peaks have ordinates respectively 6.48 (43.68), 6.46 (33.96) and 6.28 (39.22). The fourth peak is that at the frequency of 9.42 with height equal to 5.71 (Here and in the following to each line should be considered attributed an uncertainty of 0.09, as stemming from the FWHM of the spectral lines). This last peak essentially coincides with the line at 9.4 emerging from the investigation of the 10 day data set, while the other three did not play any role in the analysis of that dataset. From a Monte Carlo similar to that used to derive the Fig. 5 we get the results reported in Fig. 26. In particular, it is derived that the distribution of the highest peak is fit to the model function with $M = 639$. The distributions of the other three less amplitude peaks are somehow also described by the same M value, even if individually they would be better fit through a slightly different (lower) M value. Essentially it is reproduced the same situation of the Lomb Scargle analysis of the 10 day dataset, with the null hypothesis distributions following reasonably well the model functions.

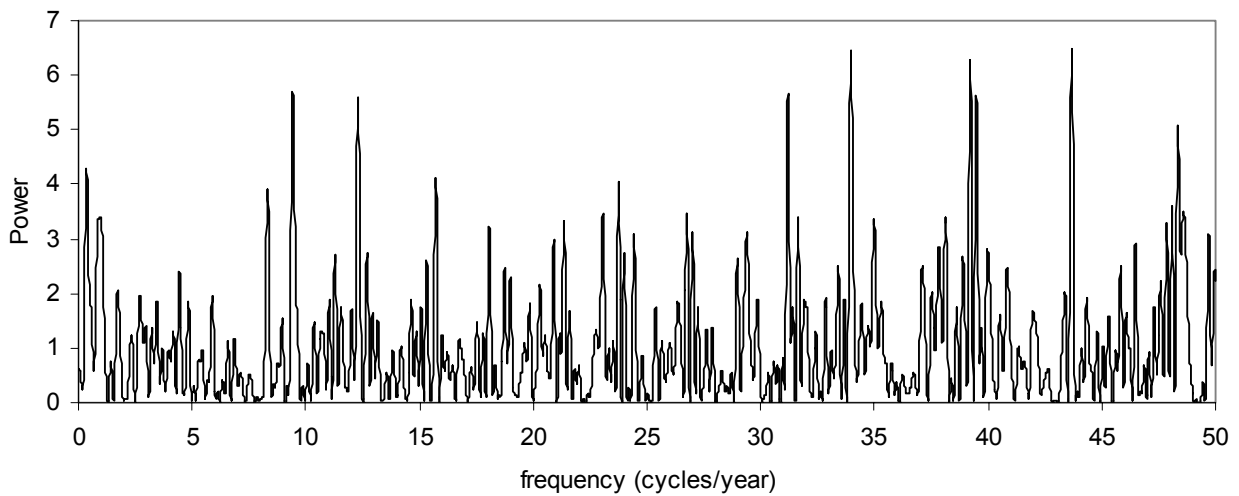

*Fig. 25 – Lomb Scargle periodogram of the 5 day binned SK data*

Quantitatively, from the Monte Carlo histograms we get that for the peak with ordinate 6.57 the significance is 59.7 % (5972 entries above 6.57 out of 10000 simulated events), that for the peak with ordinate 6.55 the significance is 21.6% (2161 entries above the 6.55 ordinate out of 10000 simulated events), that for the peak with ordinate 6.37 the significance is 8.1% (810 entries above the 6.37 ordinate out of 10000 simulated events), and that for the peak with ordinate 5.79 the significance is 9.9% (989 entries above the 5.79 ordinate out of 10000 simulated events). So, in summary, the Lomb Scargle periodogram of the 5 day binned data is consistent with a noisy series with no periodicity embedded. It is, thus, obtained the same conclusion got while analyzing the 10 day dataset with the same Lomb Scargle methodology.

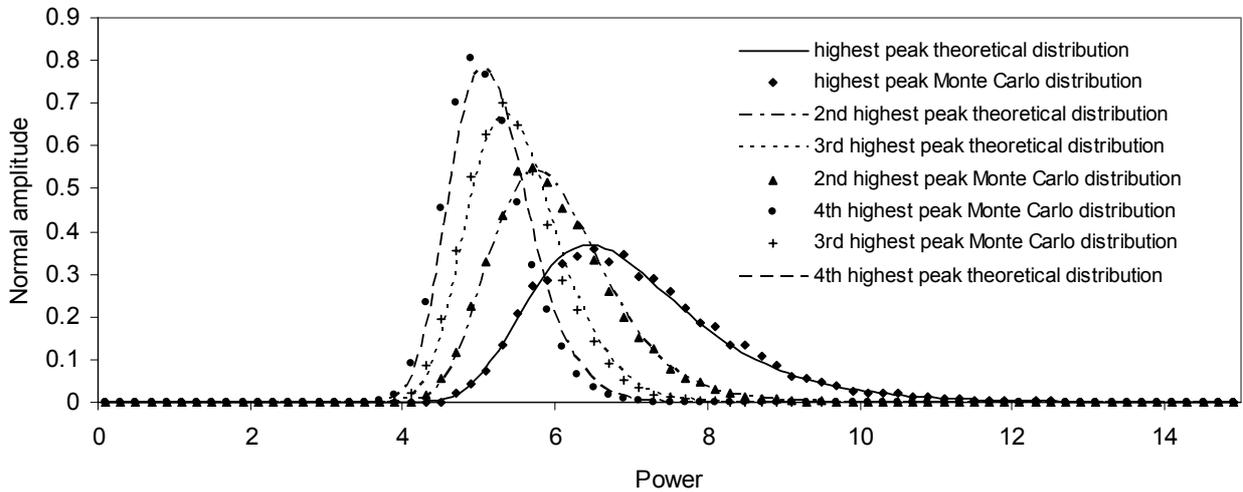

*Fig. 26 – Lomb Scargle periodogram of the 5 day binned SK data: comparison of the null hypothesis Monte Carlo distributions with the model*

### B. 5 Day Data Weighted Periodogram

The weighted periodogram of the 5 day binned SK data is shown in Fig. 27, in the frequency range from 0 to 50 cycles/year.

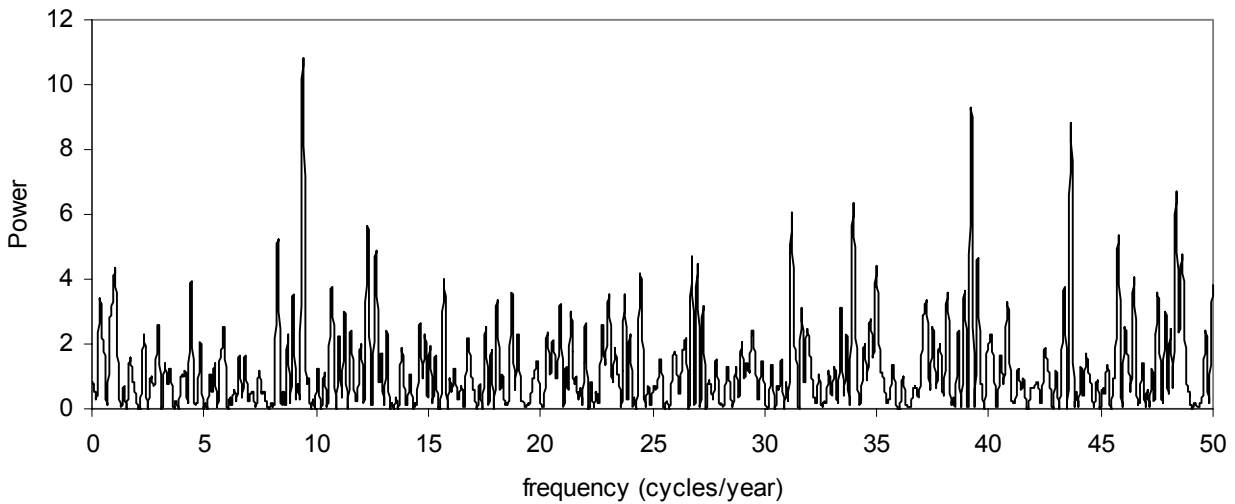

*Fig. 27 – Weighted periodogram of the 5 day binned SK data*

Now the four highest peaks have ordinates respectively 10.844 (9.42), 9.28 (39.22), 8.82 (43.66) and 6.69 (48.35). Three of them are already among the four highest also in the previous spectrum, while that at 48.35 was not present in the previous. The line at 9.42, that before was the fourth in term of height, is now the highest.

From a Monte Carlo similar to that used to derive the Fig. 6 we get the results reported in Fig. 28. It can be seen, as in the example of Fig. 6, that the Monte Carlo histograms do not follow the model functions, reported in the figure for the same value M=639 derived above. However, the tails of the four distributions are not much difference from the respective model functions, and hence are similar to the tails of the distributions pertinent to the previous Lomb Scargle case.

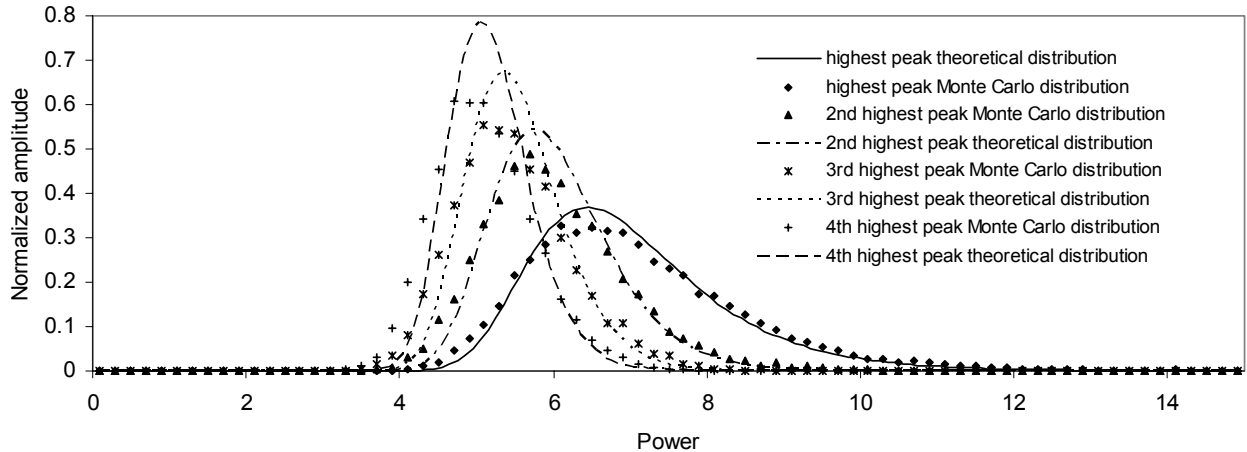

*Fig. 28 – Weighted periodogram of the 5 day binned SK data: comparison of the null hypothesis Monte Carlo distributions with the model*

Clearly, all the four highest peaks lie on the tails of the respective distributions; quantitatively, from the Monte Carlo histograms we get that for the peak with ordinate 10.844 the significance is 1.8 % (186 entries above 10.844 out of 10000 simulated events), that for the peak with ordinate 9.28 the significance is 0.5 % (54 entries above the 9.28 ordinate out of 10000 simulated events), that for the peak with ordinate 8.82 the significance is 0.1% (9 entries above the 8.82 ordinate out of 10000 simulated events), and that for the peak with ordinate 6.69 the significance is 1.8% (176 entries above the 6.69 ordinate out of 10000 simulated events). So, in summary, the weighted periodogram of the 5 day binned data appears to be not consistent with a noisy series with no periodicity embedded.

The indication got here is different from that obtained in the similar case of the 10 day spectrum, the reason being twofold: on one hand the four highest peaks exhibit an increase more pronounced now, with respect to the 10 day dataset, in passing from the Lomb Scargle to the weighted periodogram, and on the other the tails of the Monte Carlo null hypothesis distribution change toward higher values less of what happened in the 10 day analysis.

### C. 5 Day Data Likelihood Spectrum With Asymmetric Errors

The likelihood spectrum evaluated taking into account the asymmetric errors of the 5 day binned SK data is shown in Fig. 29 in the usual frequency range from 0 to 50 cycles/year. The four highest ordinates are, respectively, 9.22 (9.42), 8.97 (39.22), 8.23 (43.66) and 6.55 (31.21), hence the first three are the same of the previous spectrum. From a Monte Carlo similar to that used to derive the Fig. 7 we get the results reported in Fig. 30, which look significantly different from those related to the weighted periodogram case. It can be seen indeed that the Monte Carlo histograms deviate from the model functions plotted for the same value M =639 inferred above, with in particular the tails noteworthy enhanced .

Quantitatively, from the Monte Carlo null hypothesis distributions we get that for the peak with ordinate 9.22 the significance is 12.1 % (1208 entries above 9.22 out of 10000 simulated events), that for the peak with ordinate 8.97 the significance is 1.6% (155 entries above the 8.97 ordinate out of 10000 simulated events), that for the peak with ordinate 8.23 the significance is 0.8% (78 entries above the 8.23 ordinate out of 10000 simulated events), and that for the peak with ordinate 6.55 the significance is 6.3% (632 entries above the 6.55 ordinate out of 10000 simulated events).

So, in summary, what happens is that the trend toward stronger significance values, observed in the weighted periodogram case, is attenuated here because of the concurrent effect of the lower ordinates of the highest peaks and of the increased tails of the null hypothesis distributions.

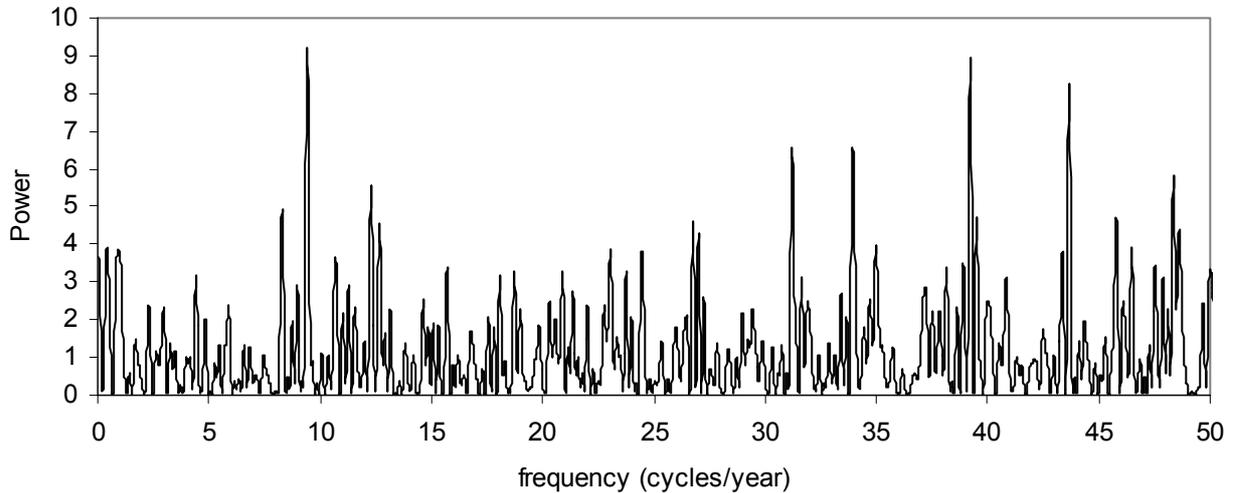

*Fig. 29 – Likelihood spectrum with asymmetric errors of the 5 day binned SK data*

From this fact it stems a controversial situation, since the overall spectrum appears only barely compatible with the constant rate hypothesis, mainly because of the significances of the second, third and fourth peak, while the significance of the highest peak considered alone would be consistent with the constant rate hypothesis. It must be added that, as in the 10 day case, the asymmetric likelihood spectrum is more similar to the weighted periodogram than to the Lomb-Scargle periodogram, and that in both spectra the highest peak is the 9.42 line, which on the other hand was only the fourth, in term of height, in the Lomb-Scargle case.

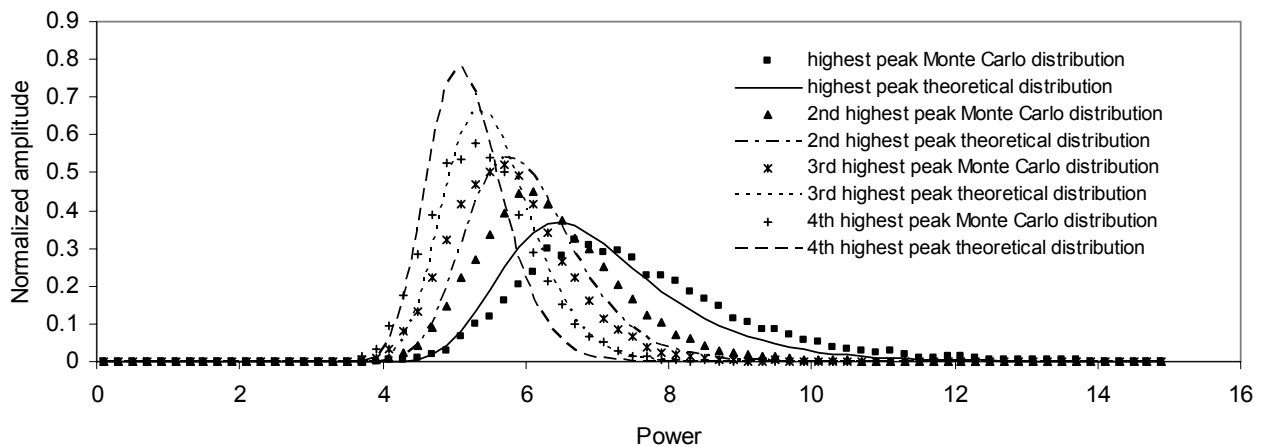

*Fig. 30 – Likelihood spectrum of the 5 day binned SK data: comparison of the null hypothesis Monte Carlo distributions with the model*

### XIII. ANALYSIS OF THE SK 5 DAY BINNED DATASET: PREDICTION OF THE SPECTRUM BEHAVIOUR IN PRESENCE OF A TRUE PERIODICITY

Let's consider again the model (62) and check its predictions related to the 5 day bin case. In Fig. 31 it is reported, over the whole search band, the expected average value of the spectral height distribution induced by a true periodicity; the plots refer to an amplitude either of 0.05 or 0.07 and to the Lomb Scargle and weighted periodogram cases; few dots are added to show also the behaviour of the likelihood spectrum with asymmetric errors. The Nyquist limit is signalled by the irregularity present on all the curves; it can be noted, as in the 10 day case, the large sensitivity to small variation of the signal amplitude, and the significant increase for the same amplitude of the average spectral response in passing from the first to the second method. This last circumstance is different from what observed in the 10 day case, where the same increase was calculated to be more modest.

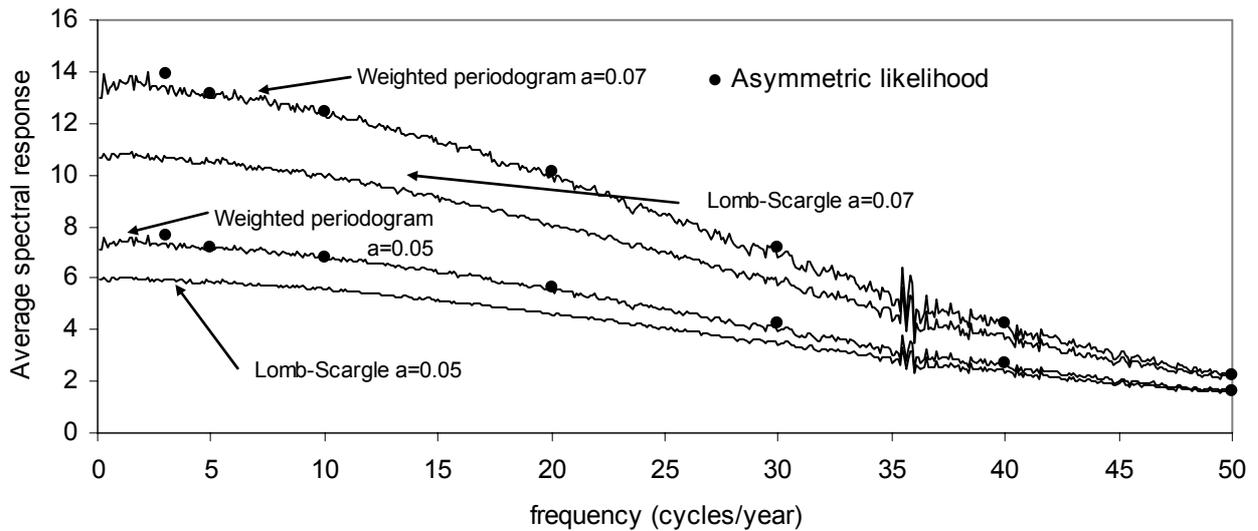

*Fig. 31 –Average spectral response over the whole search band*

Regarding the alias, in Fig. 32 it is displayed the predicted alias location for a true periodicity at the frequency of 9.42, plotted up to the extended range of 72 cycles/year: a small alias peak is predicted at the frequency of 62.55, implying that the effective Nyquist frequency is evaluated to be (Nq-9.42)+Nq=62.55, so that Nq=(62.55+9.42)/2=35.985, very close to the nominal value of 36; the alias is accompanied by an effect of power spread over contiguous frequencies that in spectrum analysis is referred as sidelobe leakage.

### XIV. SYSTEMATIC UNCERTAINTIES IN THE ERRORS HANDLING

Clearly in the above analysis a crucial role is played by the measurements errors, since it is their inclusion that produces different results with respect to those stemming in the standard Lomb-Scargle method, where the errors are simply ignored. It can thus be important to evaluate somehow the systematic effect associated with the way in which the errors are handled. A first indication of this systematic uncertainty may be already considered the difference in the significances of the highest spectral peaks between the weighted periodogram and the asymmetric likelihood methods.

A further indication of such a systematic phenomenon can be obtained by applying again the analysis of the asymmetric errors, but inverting the inequality signs in the prescriptions associated with the (22). Such an inversion would mean that one considers the Gaussian curves described by the errors as centered on the presumed values instead on the measured values. Very interestingly, the difference in the results in the significance calculation are striking; for the 5 day bin dataset one would obtain that the four highest spectra peaks would be (as usual in parenthesis there is the frequency values) 13.17 (9.42), 10.02 (39.22), 9.85 (43.66) and 7.98 (48.35). From the Monte Carlo analysis we would get that for the peak with ordinate 13.17 the significance is 0.2 % (21 entries above 13.17 out of 10000 simulated events), that for the peak with ordinate 10.02 the significance is 0.1% (8 entries above the 10.02 ordinate out of 10000 simulated events), that for the peak with ordinate 9.85 the significance is better than 0.01% (no entries above the 9.85 ordinate out of 10000 simulated events), and that for the peak with ordinate 7.98 the significance is 0.1% (11 entries above the 7.98 ordinate out of 10000 simulated events).

So, comparing these results with those obtained in the previous paragraphs it can be inferred that in all the evaluations taking account the errors the first three highest peaks are the same (the fourth being different), in particular with the 9.42 peak being always the highest, while the significances can change drastically. While it is reassuring that the same frequency is singled out in the analysis as the most prominent, independently from the errors handling, the large variability in the significance values implies that the implicit assumption in the weighted periodogram analysis that averaging the errors would not change much the significance results is not true. It is thus appropriate to give more credibility to the significance analysis of the asymmetric likelihood method, in which the errors are taken as they are; this is the attitude taken in the next paragraph, where a comprehensive data-model comparison will be attempted.

## XV. OVERALL COMPARISON OF THE MODEL PREDICTIONS WITH THE DATA

It is presented here an overall interpretation of the detected behaviour of the 5 day SK spectra, in the light of the results of paragraphs 12, 13 and 14 obtained with the various methods of analysis, integrated also with the results stemming from the investigation performed on the 10 day dataset.

Prior to go ahead with such a discussion, let's point out that the fit procedure associated with the different analyses gives for the line at 9.42 a relative amplitude of 0.056±0.016 in the case of the Lomb-Scargle analysis, 0.066±0.014 in the framework of the weighted periodogran methodology and $0.062 \pm^{0.015}_{0.014}$ for the likelihood with asymmetric errors.

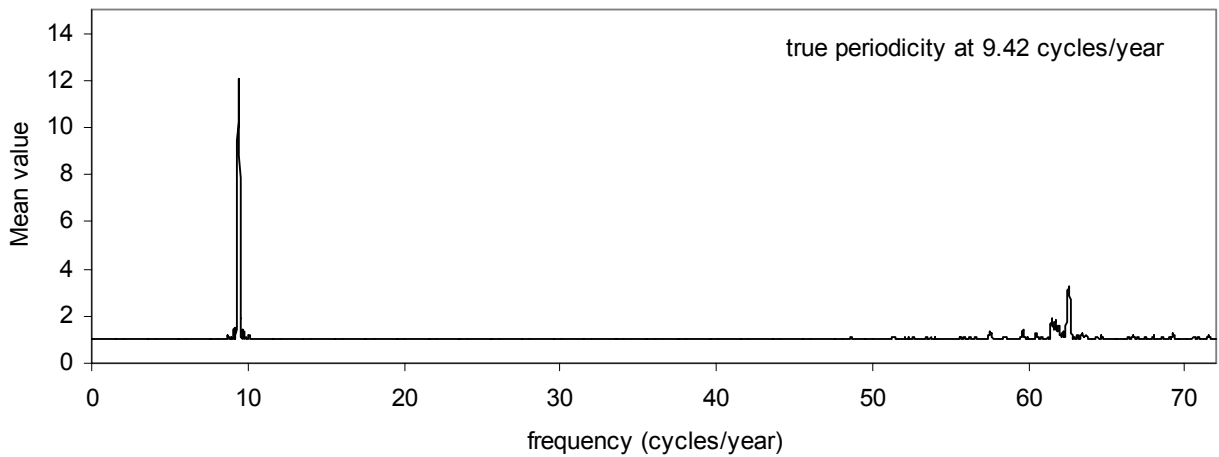

*Fig. 32 – Predicted alias frequencies of a real modulation with frequency equal to 9.42 cycles/year*

The evaluation of the potential effect thus is only slightly larger than in the 10 day case, and compatible within the uncertainties.

The summary of the comparison with the null hypothesis distributions is the confirmation of the same tendency present in the 10 day spectra: in the weighted periodogram case a significant enhancement of the ordinates of the more prominent peaks, contrasted by a modest modification of the tails of the Monte Carlo distributions with the consequence of stronger significance values with respect to the Lomb-Scargle case; in the asymmetric likelihood case a less pronounced increase of the peaks ordinates, compared with enhanced tails in the Monte Carlo distributions, leading to significances which are still stronger than the Lomb-Scargle but less than the weighted periodogram case. Furthermore, the peaks increase phenomenon is more pronounced with respect to the 10 day dataset; as net effect, the resulting weighted periodogram spectrum is in strong disagreement with the hypothesis of constant rate (but we know that in terms of significance the results of the weighted periodogram method are questionable), while the situation is more complex for the correct asymmetric likelihood spectrum, for the highest peak alone would be still compatible with a pure constant rate hypothesis, whilst the multiple peak analysis would seem at odd with such a conclusion.

In the comparison of the spectra with the predicted spectral response reported in Fig. 31 it can be noted that, as in the 10 day case, three of the four more prominent peaks, those at frequency 39.22, 43.66 and 48.35 (31.21 in the asymmetric case), fall beyond the Nyquist limit in a region of decreased sensitivity of the spectral analysis, while that at 9.42 is well within the region of maximum sensitivity.

Focusing the attention to this line (which is the only peak presents prominently both in the 5 day and 10 day spectra) it must be noted the increase of its ordinate of about 90% in passing from the Lomb Scargle to the weighted periodogram case, and of about 61% from the Lomb Scargle to the likelihood with asymmetric errors. These numbers can be compared with the predicted average increase as inferred from the model, which is of the order of 27% (a little less in the weighted periodogram case and slightly more in the asymmetric likelihood configuration). So, the model actually predicts an enhancement of the ordinate of a spectral line corresponding to a true periodicity, but much less than the amount observed in the data. Qualitatively, this is the same situation observed in the 10 day bin spectra, with the difference that here both the model increase and the data increase are amplified; similarly to the 10 day case, the data-model disagreement is less pronounced in the asymmetric spectrum.

In any case, this 27% increase prediction in the spectral ordinate corresponding to a true periodicity, together with the small modification of the null hypothesis distributions across the three methods (such a modification is more evident in the third case), demonstrate that the 5 day dataset is in principle more sensitive than the 10 day dataset in unravelling real signals embedded in the time series, via the comparison of the spectra computed through the three approaches under consideration: indeed what is expected on the basis of the model is an enhancement of the signal line(s) above a pretty constant "floor" of noise peaks. In this framework, hence, the more marked progression toward higher values of the 9.42 spectral ordinate in the 5 day case with respect to the 10 day one, could be well considered the practical manifestation of such enhanced sensitivity. However, in the context of each dataset taken separately, the quantitative disagreement between the detected and expected increase of the height of the suspect 9.42 line is surely large.

For the sake of completeness it must be added that this data-model inconsistency seems to be manifested also in the "anomalous" increase of some low level lines. As said above, actually the null hypothesis distributions exhibit a tendency to feature longer tails in the second and especially in the third methods, but to an amount that seems not adequate to describe the enhancement of some low amplitudes lines. An example is a line at 12.7, which passes from an height of 2.74 to 4.86 in the transition from the Lomb-Scargle to the weighted periodogram. So, its absolute value is such that it does not emerge from the noise "floor" in neither cases, but the percentage increase is anyhow remarkable.

Another important aspect of the present discussion is, obviously, the alias effect. In the 10 day spectrum the analytical model showed that a true periodicity at 9.4 should have been accompanied by an alias at 26.55, practically coincident with the line at 26.52 found in the spectra. Now, in the 5 day spectra there is no more hint of this second line, and so the prediction of the alias calculation showed in Fig. 32 is very interesting, since even there the line at 26.55 actually is no more present.

Regarding the small peak at 62.55, which on the other hand is computed by the alias model as displayed in Fig. 32, it could be useful to check if there is any hint of it in the actual spectra. This is done in Fig. 33 and 34 where there are the two alias predictions, in the expanded spectral region between 60 and 65 cycles/year, for the Lomb-Scargle periodogram (computed assuming for the line 9.42 the amplitude of 0.056 mentioned at the beginning of this paragraph) and for the weighted periodogram (computed assuming an amplitude of 0.066). Overlapped to the prediction, the respective spectra are displayed, too, included in the second plot the asymmetric likelihood spectrum.

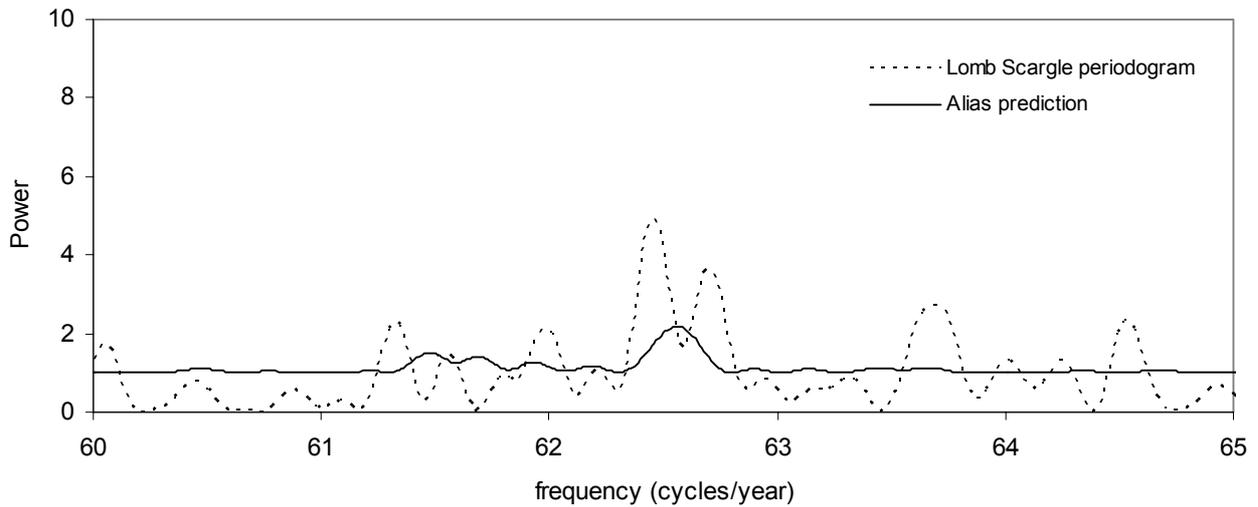

*Fig. 33 – 5 day Lomb-Scargle spectrum: comparison of the alias prediction and the actual spectrum in the expected alias region*

From both figures it is inferred that the expected alias effect is not a sharp peak but a bump. The expected average value of such a bump is higher in Fig. 34 than in Fig. 33. In the actual spectra it is observed in correspondence of the alias bump two peaks somehow merged together; the merging is more pronounced in the latter figure (weighted periodogram and asymmetric likelihood), and less evident in the former (Lomb-Scargle periodogram).

Such a behaviour is consistent with the presence of the bump predicted by the alias model, in the sense that the two peaks are likely to be two noise peaks, with the bump just located to fill the valley between them, more in the latter spectra and less in the former spectrum. Even if the effect is small, the prediction of the model is very distinctive and peculiar, and matches the feature of the experimental spectra. Clearly, the simple comparison of the alias model with the standard Lomb-Scargle spectrum alone would have not given much insight in this phenomenon, while the successive comparison of the model with all the three spectra proves to be the key to shed light on the likely presence of the effect. In any case, as already pointed out in the 10 day case, the consistency of the detected spectral behaviour with the alias prediction per se is not a proof of the presence of a real signal, due to the noise correlation effects which can emulate a similar behaviour.

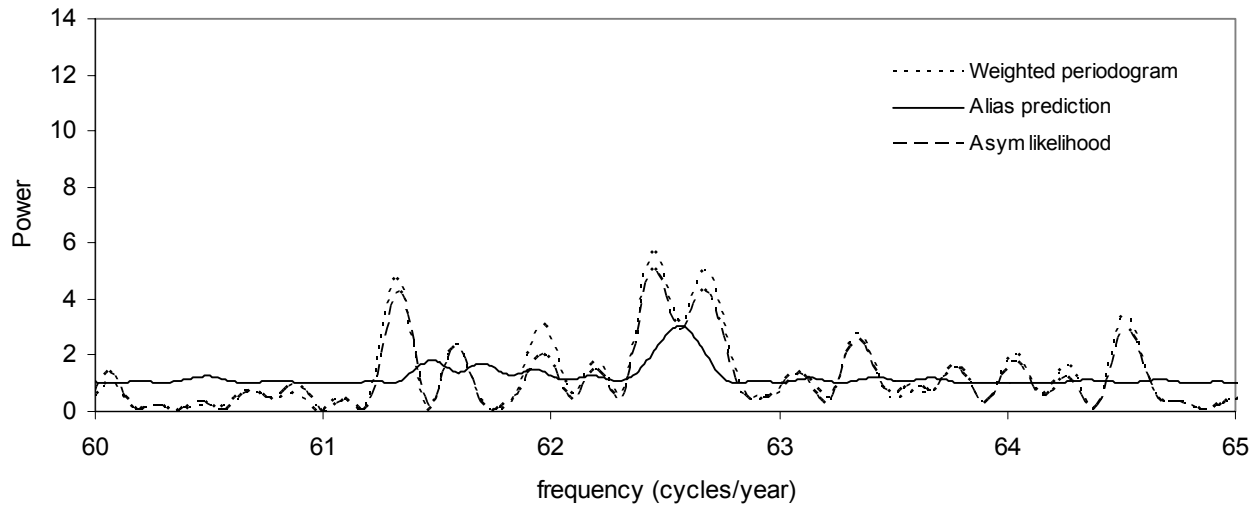

*Fig. 34 – 5 day weighted and asymmetric likelihood spectra: comparison of the alias prediction and the actual spectra in the expected alias region*

This consistency, on the other hand, could make sense in the framework of a more general set of indications concurrently pointing towards of the presence of a signal.

At this point putting together all the pieces of the puzzle, the following summary can be done:

a) the 9.42 line emerges clearly from the analysis of both datasets;

b) in the 10 day data this line is accompanied by an alias line that disappears in the 5 day dataset; both effects are in agreement with the prediction of the alias calculations. Furthermore, a distinctive alias bump is predicted by the model in the 5 day case, which matches the observed spectral behaviour;

c) both in the spectra related to the 10 and 5 day cases, the 9.42 spectral ordinate increases from the first to the second and third analysis methods, qualitative in agreement with the model but quantitatively in disagreement with it (the increase is more than expected);

d) the trend of the significance assessment values is somehow puzzling; the situation in the 10 day case referred in the previous paragraph 11 can be succinctly summarized reminding that the three original spectra are consistent with the hypothesis of constant rate, with some notes and caveats that are thoroughly described at the end of that paragraph.

On the other hand, for the 5 day dataset there is a distinctive difference considering either the significance of the 9.42 line alone or the multiple peaks significance assessment. The 9.42 line is the highest both in the second and third method, but with large difference in the significance value: in the weighted periodogram approach with the crude errors averaging the significance is as strong as 1.8%, while it is only about 12.1% in the likelihood with the correct treatment of the asymmetric errors. Hence in this framework the significance of the highest peak in the method which treats the errors correctly is a clear indication in favour of the consistency with the constant rate hypothesis. On the contrary, the concurrent assessment of the significance of the highest peaks, in both cases, does not provide further support to the constant rate hypothesis because of the low significance values, especially of the second and third peak in the likelihood spectrum.

Thus the plurality of evaluations listed in the above points a), b), c) and d), stemming from the calculations of the model for the time series spectral analysis described in the first part of this work, depicts a controversial scenario: the significance of the highest spectral peak in the correct asymmetric spectrum is such that the constant rate hypothesis cannot be excluded, but the other indications (multiple peaks significance assessment, alias, increase of the 9.42 spectral ordinate from the Lomb-Scargle method to the methods including the errors) do not seem in line with this conclusion. However, two of these potential counter-indications have to be considered with some care, for the alias can be mimicked also by the correlation of noise peaks and the ordinate increase of the 9.42 line is not in agreement with the model prediction.

Hence, the strongest counter-indication of the constant rate hypothesis appears to be the multiple peaks significance assessment. To shed further light on this point, it can be attempted the same exercise tried in the 10 day case, i.e. the highest 9.42 peak is ignored and the significance of the remaining peaks re-evaluated. In such a way it is obtained that the three highest remaining peaks at frequencies 39.22, 43.66 and 48.35 (31.21 in the asymmetric likelihood) feature, respectively, the significance values of 7.6%, 1.04% and 6.8% in the weighted periodogram spectrum, and 14.6%, 4.4% and 16.4% in the likelihood with asymmetric errors. It is, therefore, interesting to note that, ignoring the highest peak, the three prominent peaks left over would produce significance values that, in the case of the correct asymmetric likelihood, can be considered in agreement with the hypothesis of consistency with a pure noisy spectrum.

For the sake of completeness, it must be added that the peaks at 39.22 and 43.66, whose ordinates as that of the 9.42 line increases substantially in the second and third method if compared with the Lomb-Scargle case, appear also in the 10 day case, even though with reduced height.

In summary, the significance of the highest peak at 9.42 cycles/year in the correct asymmetric likelihood spectrum would lead to conclude that the consistency of the SK data with the constant rate hypothesis cannot be excluded, but the other contradictory indications which do not fully support such consistency point towards the desirability of further investigations of the same dataset through the exploitation of different analysis techniques.

As conclusive remark of this paragraph it should be added that the plausible data-model inconsistency that has been detected means that presumably we are scanning the data with an un-appropriate "instrument", i.e. the Gaussian assumption. Recently a number of evaluation of time series behaviour in the presence of non Gaussian tailed noise has been published, see for example [21], showing that deviations from gaussianity have significant impacts on the spectrum statistical properties. Since in [1] it is explained that the data uncertainties have been estimated by asymmetric Gaussian approximation of the unbinned maximum likelihood fit to the $\cos\theta_{sun}$ distributions, the investigation of the precise errors tail shape and of its effect in the analysis could be a clue towards the understanding of the model-data disagreement.

## XVI. DETECTION EFFICIENCY

It can be useful to conclude this work with a digression on the detection efficiency implied by a sampling scheme and errors like that featured by the 5 day binned SK data. Such a discussion explains also quantitatively why an effect in the 5-7% range is difficult to catch.

It is well known that the signal-to-noise discrimination in repeated experiments is usually quantified with the probability of false alarm and the corresponding probability of detecting a signal if actually present, i.e. given the noise distribution a threshold is defined which ensures the desired false alarm level, and consequently the integral above that threshold of the signal induced distribution gives the respective detection probability. This procedure is of special use in real time laboratory experiments, when the threshold is set through a discriminator so to allow the acquisition only of the signals of interest; indeed having defined a priori via a calibration the threshold level corresponding typically to a few percent of false alarm value, the experimentalist knows how pure the acquired sample of events is. This way of presenting things is clearly not the best when you have only one event to judge. In the present case the single event is the spectrum of the process, on which it would be very limiting to perform only a yes/no decision inference: what it seems more appropriate, instead, is to perform on it a plurality of investigations, as shown in the previous paragraphs.

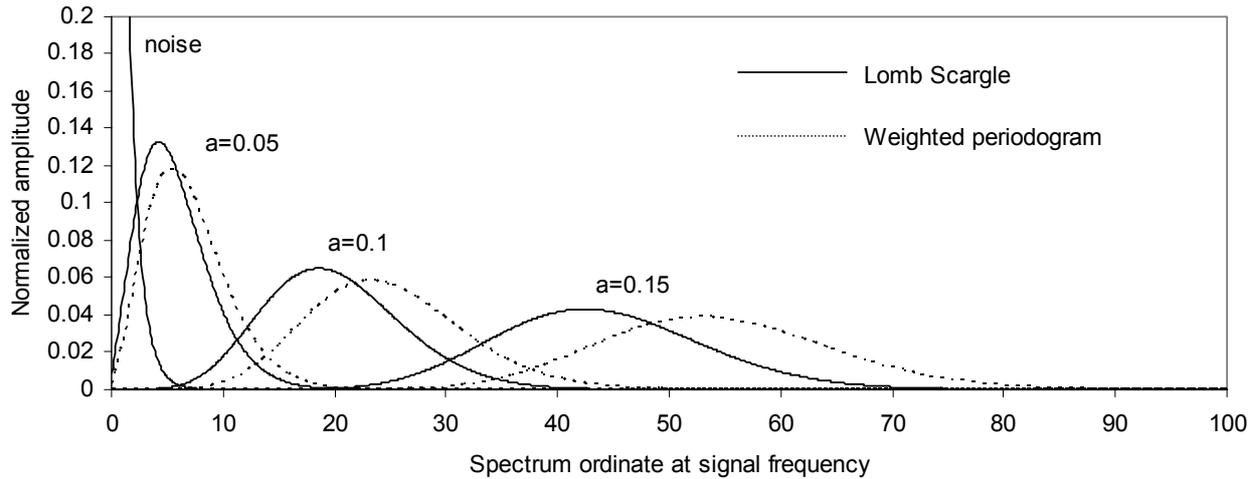

*Fig. 35 – Comparison of the spectrum ordinate distributions at signal frequency related to the Lomb Scargle and weighted periodogram methods corresponding to different oscillation amplitude, overlapped to the single frequency noise distribution*

Despite this caveat, it is of some interest, also for the purpose of comparison with published results, to imagine the experiment repeated many times and calculate the expected detection efficiencies for a prescribed false alarm level. The ingredients to perform such an evaluation are the actual noise and signal distributions. Since we are dealing with an hypothetical, standardized yes/no procedure, the signal to noise comparison can be only the simple, straightforward comparison with the highest noise spectral peak distribution.

It must be kept in mind that the scenario is much different if we are searching for a signal at a pre-defined frequency or if we perform a blind scan over a search band. The former case, assuming a modulation at the example frequency of 5 cycles/year, is depicted in Fig. 35: the distribution of the noise as thoroughly explained in paragraph 7 is a simple exponential curve, while the distribution related to the signal is the non central chi square distribution (62), that is reported in the figure for the three relative amplitudes of 0.05, 0.1 and 0.15, both for the Lomb Scargle (solid lines) and weighted periodogram (dotted lines) methodologies (the Monte Carlo curves for the likelihood with asymmetric errors are not reported since they differ only slightly from those stemming from the application of the weighted periodogram).

Since the curves related to the weighted periodogram approach are systematically shifted towards higher values with respect to those pertaining to the application of the standard Lomb Scargle periodogram, while the noise is exponential distributed for both cases, it stems that the degree of overlap of the noise and signal distributions is higher in the latter case than in the former; this fact implies that the weighted periodogram methodology features an higher detection efficiency than the Lomb Scargle method.

The more interesting occurrence is, however, that in which the frequency of the signal is unknown and hence the spectrum scanned over a search band. The Fig. 35 must be accordingly modified, in the sense that now the noise distribution to be considered is the amplitude distribution of the largest noise spectral peak, as extensively determined via the previous Monte Carlo evaluations. The situation is, however, somehow qualitatively similar because the Monte Carlo showed that the largest peak distribution do not differ much among the various analysis methods. For illustrative purpose, in Fig. 36 the single frequency noise exponential distribution has been replaced with the largest spectral noise peak distribution of Fig. 28, evaluated for the weighted periodogram case: clearly the noise distribution, being shifted toward higher values, now hinders severely the detection of small amplitude oscillations.

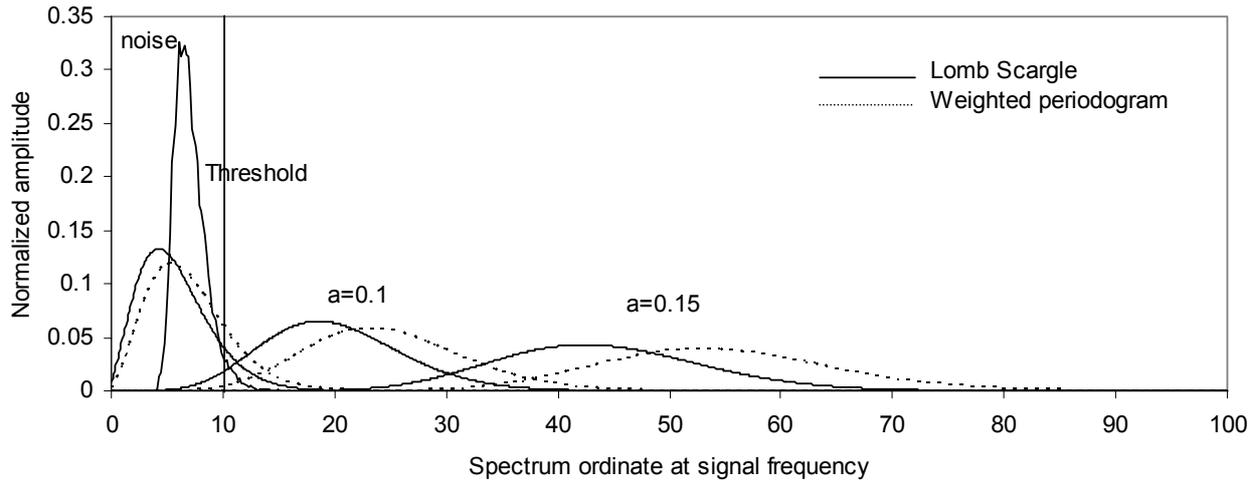

*Fig. 36 – Comparison of the spectrum ordinate distributions at signal frequency related to the Lomb Scargle and weighted periodogram methods corresponding to different oscillation amplitude, overlapped to the whole frequency search band noise distribution*

The plot of Fig. 36 can be transformed into an expressive series of iso-sensitivity contours on a frequency-amplitude plane. For this purpose one sets the maximum acceptable false alarm probability, for example 98%, and from the Monte Carlo noise distribution derives the corresponding threshold (vertical line in the figure). The integral above this limit of the signal distribution corresponding to a certain signal amplitude and frequency gives the relevant detection efficiency at the predefined false alarm probability level. Repeating many times such a computation it is possible to construct the iso-sensitivity contours shown in Fig. 37 and 38.

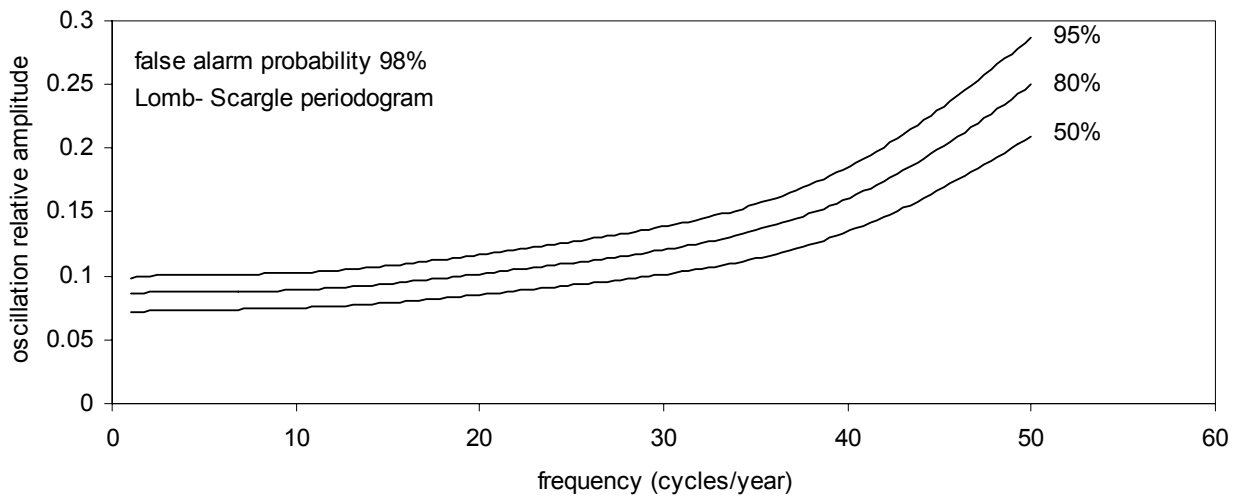

*Fig. 37 – Iso-sensitivity contours for the Lomb-Scargle method for the prescribed false alarm probability of 98%*

The contours in the figures represent the loci on the amplitude frequency plane that correspond to the 95%, 80% and 50% detection efficiency, for the given false alarm probability of 98%. Those in Fig 37 are related to the application of the standard Lomb Scargle periodogram, while those in Fig. 38 result from the use of the weighted periodogram; in this latter plot few dots have been added to check the sensitivity also of the asymmetric likelihood method: the dots are shifted slightly above the corresponding continuous curves, indicating a modest sensitivity reduction in the asymmetric spectrum case.

By comparing the two figures it can be inferred that the increase of sensitivity in passing from the former to the latter method is of the order of 12 %: for example while the sensitivity contour of the Lomb Scargle method, for the 95% detection efficiency and at low frequency, corresponds to an amplitude of oscillation of 10%, the sensitivity contour for the weighted periodogram for the same detection efficiency at low frequency corresponds to an amplitude of 8.8%. The two figures also demonstrate that the increase of sensitivity from the former to the latter method is systematic over the whole frequency range.

It must be pointed out that the 95% sensitivity contour for the 98% false alarm probability and for the Lomb Scargle method was also computed by the Super-Kamiokande collaboration in [1]. The contour they reported in the Fig. 7 of that reference, apart a display difference due to the difference choice of the variable on the x axis (period instead of frequency), coincides well with the corresponding contour reported here in the Fig 37.

As final remark it should be added that from the sensitivity contours curves it can be inferred that a 7% oscillation would be detected, according to the 98% false alarm probability threshold, about 50% of the times in a sequence of hypothetical repeated experiments, if using the weighted periodogram as analysis method; clearly, keeping in mind the discussion of paragraph 15, this prediction is error model dependent (as it stands, it is valid for Gaussian errors) and may thus change accordingly.

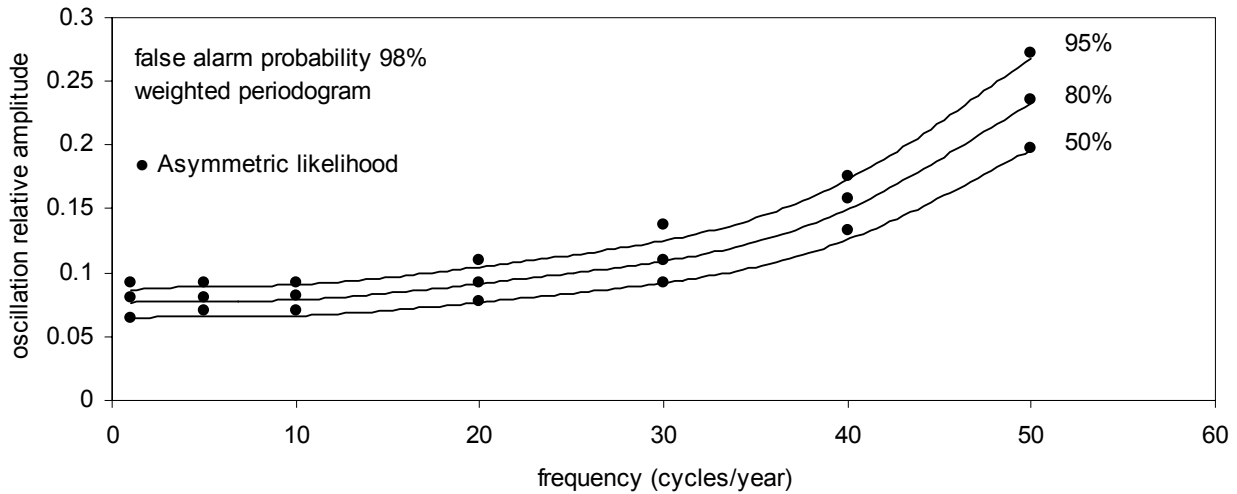

*Fig. 38 – Iso-sensitivity contours for the weighted periodogram method for the prescribed false alarm probability of 98%*

## XVII. SUMMARY

In summary this work is constituted by two main parts.

The first part is focused to a very thorough review of the analysis methods present in the literature for the periodogram-like spectral analysis of time series. While doing this review it has been pointed out the intimate relationship between the standard Lomb-Scargle methodology and the more general likelihood approach, of which the Lomb-Scargle periodogram is simply the special case in which the errors are not considered. It has been also illustrated how the likelihood method allows to take into account the errors, even if asymmetric and without a prior averaging.

This first part includes also a deep illustration of the spectrum statistical properties, both in the null hypothesis and in the case of a true periodicity embedded in the series: two important points in this context are the introduction of a model describing the probability density functions for all the subsequent spectral peaks, and not only for the highest as usually done in the literature, and the description of the model giving the probability density function of the spectrum ordinate corresponding to a true periodicity, which can be also powerfully used to predict the associated alias effects. The first part is concluded with some validation examples, including careful comparisons of the above models with Monte Carlo calculations.

The second part of this work is devoted to apply the mathematical tools shaped in the first part to the Super-Kamiokande solar neutrino time series data, both the 10 day and 5 day binned datasets. Succinctly, what stems from the calculation is the following: the standard Lomb Scargle methodology, which is a subset of the previous model using only a part of its statistical capabilities, does not reveal hints of periodicity embedded in the data, in agreement with the Super-Kamiokande analysis. On the other hand, the inclusion in the analysis of all the statistical tools of the model, i.e. errors both in the symmetric or asymmetric form, multiple peaks significance assessment and alias prediction, delineates a more complex picture in which a line at 9.42 cycles/year (a periodicity thus of almost 39 days, already pointed out in the various analyses of Sturrock and collaborators) emerges in the spectrum with an individual significance which cannot exclude the constant rate hypothesis, but accompanied by other indicators that do not fully endorse such a conclusion. Therefore, in order to shed further light on the features of the SK time series data, and in particular on the mentioned discrepancy, more independent analyses of the same dataset would be desirable, possibly carried out exploiting alternative investigation techniques.

## AKNOWLEDGMENTS

The author would like to thank Raju Raghavan for many useful discussions, Masayuki Nakahata for an intense and fruitful exchange of communications, the Super-Kamiokande Collaboration for making the data publicly available.

## REFERENCES

[1] J.Yoo et al., "Search for periodic modulations of the solar neutrino flux in Super-Kamiokande-I", Physical Review D, vol. 68, Issue 9, November 2003, id. 092002

[2] D.O. Caldwell and P.A. Sturrock., "Evidence for solar neutrino flux variability and its implications", Astroparticle physics, vol. 23, (2005), 543-556

[3] P. A. Sturrock et al., "Power-spectrum analyses of Super-Kamiokande solar neutrino data: variability and its implications for solar physics and neutrino physics ", hep-ph/0501205

[4] G. Ranucci, "Time series analysis methods applied to the Super-Kamiokande I data ", hep-ph/0505062

[5] N.R. Lomb, "Least-squares frequency analysis of unequally spaced data", Astrophysics and Space Science, vol. 39, Feb. 1976, p. 447-462

[6] J.D. Scargle, "Studies in astronomical time series analysis. II - Statistical aspects of spectral analysis of unevenly spaced data", Astrophysical Journal, Part 1, vol. 263, Dec. 15, 1982, p. 835-853

[7] E. J.,Groth, " Probability distributions related to power spectra", Astrophysical Journal Supp. Ser. 29, (1975), 285-302

[8] F.J.M. Barning, "The numerical analysis of the light-curve of 12 Lacertae ", Bull. Astron. Inst. Neth., Volume 17, no. 1, pp 22-28 (1963)

[9] J.D. Scargle, "Studies in astronomical time series analysis. III – Fourier Transforms, autocorrelation functions, and cross-correlation functions of unevenly spaced data ", Astrophysical Journal, vol. 343, Aug. 15, 1989, p. 874-887

[10] A. Schuster, "On the investigation of hidden periodicities with application to a supposed 26 day period of meteorological phenomena", Terrestrial Magnetism, 3,13-41, 1898

[11] S.S. Wilks, "The large-sample distribution of the likelihood ration for testing composite hypotheses ", Ann. Math. Stat., 9, (1938) 60-62

[12] W.T. Eadie et al. "Statistical methods in experimental physiscs", North-Holland Editions (1971)

[13] W.H. Press et al., "Numerical recipes in Fortran ", Cambridge University Press (1991), sect. 12.1

[14] W.H. Press et al., "Numerical recipes in Fortran ", Cambridge University Press (1991), sect. 13.8

[15] G. Ranucci, "Time statistics of the photoelectron emission process in scintillation counters", Nucl. Instr. and Meth. A335, 15 October 1993, p. 121-128

[16] Wang Zhaomin et al, "The influence of average photon number on the measured fluorescence decay time of scintillator ", Nucl. Instr. and Meth. A419, 11 December 1998, p. 154-159

[17] A. Papoulis, "Probability, random variables, and stochastic processes ", McGraw-Hill (1984), $2^{nd}$ edition

[18] Horne, J, H..and Baliunas, S., L., "A prescription for period analysis of unevenly sampled time series",  Astrophysical Journal, vol. 302, March 15, 1986, p. 757-763

[19] R.A. Fisher, "Tests of significance in harmonic analysis", Proceedings of the Royal Society of London, A, 125 (1929) 54-59

[20] A. F. Siegel,"Testing for periodicity in a time series", Journ. Am. Stat. Assoc., vol. 75, No 370 (1980), 345-348

[21] W. X. Zhou and D. Sornette, "Statistical significance of periodicity and log-periodicity with heavy-tailed correlated noise", International Journal of Modern Physics C  13, 137-170 (2002)